%% file: arxiv.tex
\documentclass[aps,pre,reprint,floatfix]{revtex4-1}
\usepackage{blindtext}
\usepackage{lipsum}
\usepackage{amsmath}
\usepackage{amsfonts}
\usepackage{amssymb}
\usepackage{comment}
\usepackage{graphicx}
\usepackage{dcolumn}
\usepackage{bm}
\usepackage{physics}
\usepackage[normalem]{ulem}
\usepackage{xcolor}
\usepackage{xr-hyper} 
\usepackage{hyperref} 
\usepackage{algorithm}
\usepackage{algpseudocode}
\usepackage{tcolorbox}

\newcommand{\cM}{\mathcal{M}}
\newcommand{\Real}{\mathbb{R}}

\input{commands}

\begin{document}
\title{Nonlinear classification of neural manifolds with contextual information}
\author{Francesca Mignacco$^{1,2}$, Chi-Ning Chou$^{3}$ and SueYeon Chung$^{3,4}$}
 \affiliation{$^1$Graduate Center, City University of New York, New York, NY 10016, USA\\$^2$Joseph Henry Laboratories of Physics, Princeton University, Princeton, NJ 08544, USA\\$^3$Center for Computational Neuroscience, Flatiron Institute, New York, NY 10010, USA\\$^4$Center for Neural Science, New York University, New York, NY 10003, USA}

\begin{abstract}
Understanding how neural systems efficiently process information through distributed representations is a fundamental challenge at the interface of neuroscience and machine learning. Recent approaches analyze the statistical and geometrical attributes of neural representations as population-level mechanistic descriptors of task implementation. In particular, \emph{manifold capacity} has emerged as a promising framework linking population geometry to the separability of neural manifolds. However, this metric has been limited to linear readouts. To address this limitation, we introduce a theoretical framework that leverages latent directions in input space, which can be related to contextual information. We derive an exact formula for the context-dependent manifold capacity that depends on manifold geometry and context correlations, and validate it on synthetic and real data. Our framework's increased expressivity captures representation reformatting in deep networks at early stages of the layer hierarchy, previously inaccessible to analysis. As context-dependent nonlinearity is ubiquitous in neural systems, our data-driven and theoretically grounded approach promises to elucidate context-dependent computation across scales, datasets, and models. 
\end{abstract}

\maketitle
\section{Introduction}
 Understanding the neural population code that underlies efficient representations is crucial for neuroscience and machine learning. Approaches focused on the geometry of task structures in neural population activities have recently emerged as a promising direction for understanding information processing in neural systems \cite{chung2021neural,sorscher2022neural,bernardi2020geometry,ansuini2019intrinsic}.  Notably, analytical advances linking representation geometry to the capacity of the downstream readout \cite{chung2018classification,wakhloo2023linear} have shown a promise as a normative theory and data analysis tool, providing a pathway for explicitly connecting the structure of neural representations and the amount of emergent task information \cite{cohen2020separability,kuoch2024probing}. Specifically, Ref.~\cite{chung2018classification} introduced {\em neural manifold capacity} as a measure of representation untanglement. In neuroscience, the term {\em neural manifold} refers to the collection of neural responses originating from variability in the input stimuli for a given object  (e.g., orientation, pose, scale, location, and intensity), or from the variability generated by the system (e.g., trial-to-trial variability). The capacity measures how easily random binary partitions of a set of manifolds can be separated with a hyperplane and depends on geometrical attributes of these manifolds as well as their organization in neural state space.
\subsection{Motivation}
 While manifold capacity theory has successfully analyzed datasets from both biological and artificial systems~\cite{froudarakis2020object,yao2023transformation,paraouty2023sensory,chou2024, cohen2020separability}, it has been limited to the capacity of a \emph{linear} decoder. Meanwhile, recent studies have revealed limitations in the linear capacity framework, particularly when applied to early layers of deep networks. In these layers, representations are highly entangled, and linear capacity consistently approaches its lower bound \cite{cohen2020separability}. This suggests that linear readout lacks the expressiveness needed to effectively probe changes in representational untangling at these early stages. These findings highlight the need for a more comprehensive theory—one that incorporates nonlinear decoding of neural representations. Furthermore, the existing formalism overlooks an important aspect of neural processing: the ability of neurons to selectively respond to particular object classes while maintaining tolerance to object variability \cite{DICARLO2007333}.

We propose a theoretical framework to quantify separability of neural representations based on context-dependent gating nonlinearity \cite{veness2021gated}. This mechanism, in which non-overlapping sets of units activate for specific task conditions \cite{veness2021gated, masse2018alleviating}, has gained popularity in theoretical investigations due to its tractability and expressivity \cite{saxe2022neural, li2022globally, lippl2022implicit}. Initially proposed to address catastrophic forgetting problems \cite{veness2021gated, masse2018alleviating}, context-dependent gating has been studied across multiple scales in the brain. At the microscopic level, recent studies have proposed potential biological implementations, such as dendritic gating \cite{Sezener2021.03.10.434756, london2005dendritic}. At the macroscopic level, human studies have modeled context-dependent suppression of task-irrelevant information \cite{FLESCH20221258}. 
More broadly, contextual-information processing enhances cognitive and behavioral flexibility, supporting nonlinear computation across a wide range of tasks, including decision making \cite{mante2013context, panichello2021shared,FLESCH20221258} and attention \cite{moore2003selective,buschman2007top,cukur2013attention,tavoni2021cortical,buschman2015behavior}.
Despite its theoretical motivations and biological relevance, the impact of context-dependent gating on downstream task performance given neural representations remains largely unexplored.
 \subsection{Main contributions}
 In this work, we develop a theory of non-linearly-separable classification of correlated neural manifolds. {In order to implement nonlinear decisions, the computation is distributed across linear classifiers that selectively respond to sub-regions of the representation space, akin to receptive fields. The partitioning of representation space is arbitrary, which allows to adapt our framework to various interpretations and enable diverse applications. Context-dependent representations can be readily integrated in this setting.} Our contribution is twofold. First, we extend the existing geometric theory of capacity and abstraction to non-linear readouts. Second, we apply our theory to assess the separability of representations in deep neural networks as a testbed for sensory hierarchy. Adding contexts enables more sophisticated classification rules by the readout, increasing the number of piece-wise linear components of the decision boundary. {Accordingly, the number of decision neurons required for separation introduces another axis to measure the degree of untangling of a representation.} As a result, our framework quantitatively reveals that the layer hierarchy progressively untangles representations, as evidenced by an increase in capacity even in early layers. Moreover, the highly nontrivial context-dependent capacity curves highlight significant restructuring of representations, offering insights previously inaccessible to analysis \cite{cohen2020separability}. Importantly, our method is applicable to a wide range of datasets and models, both from machine learning and neural recordings, offering a unified approach to understand context-dependent computation in distributed representations across scales.
\section{Theoretical framework}
 \begin{figure}[t!]
     \centering
\includegraphics[width=.48\textwidth]{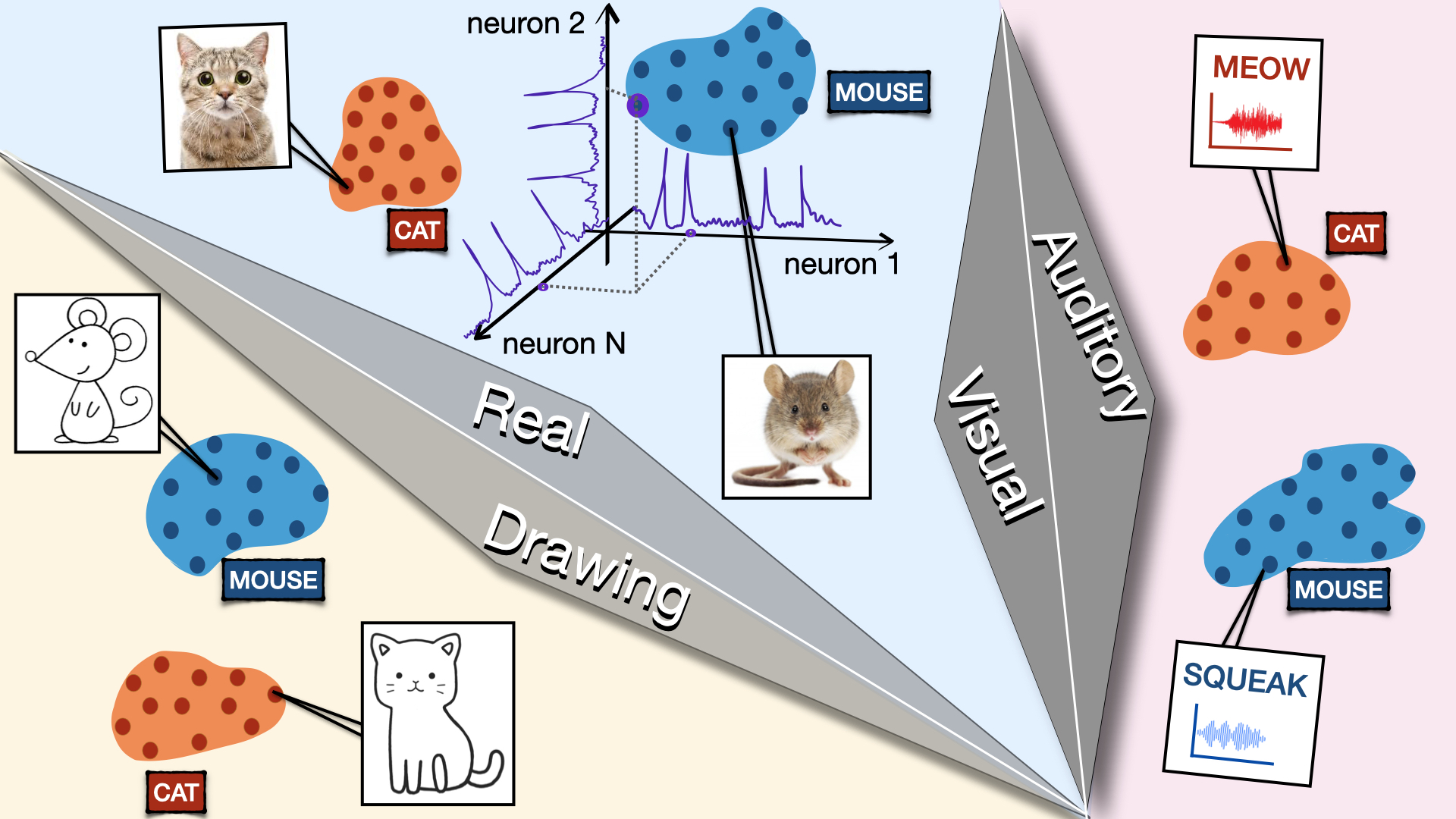}
     \caption{Neural manifolds are composed by the collection of the neural responses elicited by the same concept, either ``cat'' (red) or ``mouse'' (blue) in this illustration. These two concepts are expressed by different representations according to the different contexts: auditory or visual signals (stimulus modality as different contexts), realistic or cartoon-like images (visual styles as different contexts).}
    \label{fig:contexts1}
 \end{figure}
\subsection{The model}
\begin{figure}[t!]
\centering
\includegraphics[width=.49\textwidth]{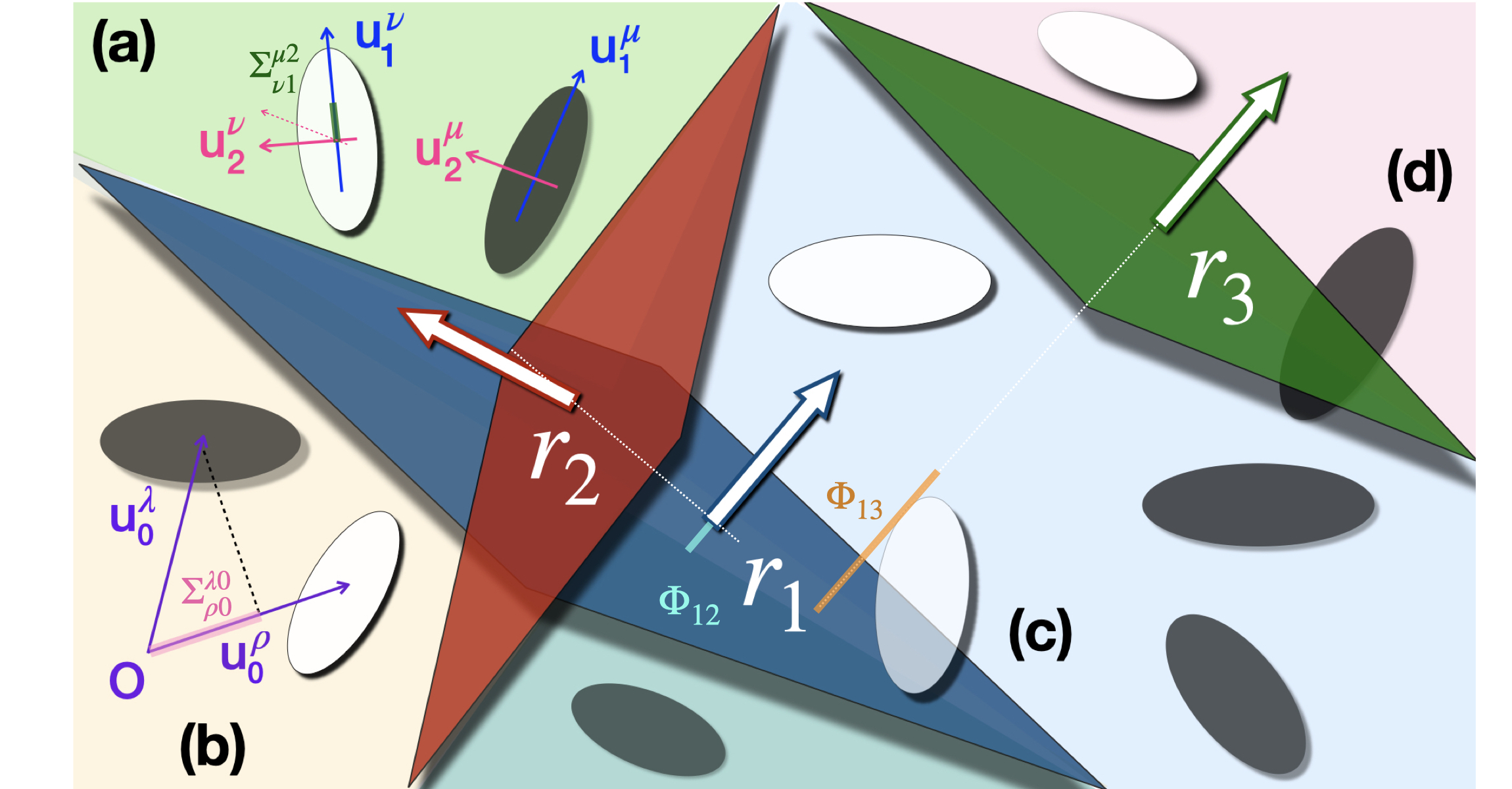}
\caption{
Three hyperplanes shatter the input space into different contexts, marked by different colors. Manifold shapes are ellipsoids, while labels are encoded by the black/white coloring. The vectors $\bm{r}_1$ and $\bm{r}_2$ define uncorrelated contexts ($\Phi_{12}\sim 0$), while $\bm{r}_1$ and $\bm{r}_3$ are highly correlated ($\Phi_{13}\sim 1$). {\bf (a)} Correlations between manifold directions. {\bf (b)} Correlation between manifold centers, where $\bf O$ denotes the origin. {\bf (c-d)} Manifolds can be ``cut'' by context hyperplanes and lie into multiple contexts. \vspace{-.3em}
\label{fig:illustration}}
\end{figure}

We model neural representations as low-dimensional manifolds embedded in high-dimensional space \cite{chung2018classification}. We consider manifolds $\mathcal{M}^\mu$, $\mu=1,\ldots,P$, each corresponding to a compact subset of an affine subspace of $\mathbb{R}^N$ with affine dimension $D\ll N$. A point on the manifold $\xx^\mu\in\mathcal{M}^\mu$ can be parametrized as
$
    \xx^\mu (\ss)=\uu^\mu_0+\sum_{i=1}^{D}s_i \uu_i^\mu$, with $\ss\in\mathcal{S}$. Here $\uu^\mu_0$ denotes the center of the manifold, while $\{\bm{u}_i^\mu\}_{i=1}^{D}$ define a $D-$dimensional linear subspace containing the manifold, and the $D$ components $s_i$ are the coordinates of the point $\xx^\mu$ on the manifold, constrained to a given shape by the set $\mathcal{S}$. We draw manifold directions $\bm{U}=\{\uu_0^\mu,\{\uu_i^\mu\}_{i=1}^D\}_{\mu=1}^P$ from the joint probability distribution
\begin{align}
\label{eq:u_distrib}
   p(\bm{U})\propto \exp\left(-\frac{N}2\sum_{\mu,\nu,i,j,l}(\bm{\Sigma}^{-1})^{\mu i}_{\nu j}u^\mu_{il}u^\nu_{jl}\right)\;,
\end{align}
where correlations between directions are encoded in the tensor $\bm{\Sigma}\in \mathbb{R}^{P(D+1)\times P(D+1)}$, as in \cite{wakhloo2023linear}.
All points on a given manifold share the same label. Labels are binary and randomly assigned: $y^\mu=\pm 1$ with equal probability. \\\indent 
At variance with \cite{chung2018classification,wakhloo2023linear}, we incorporate non-linear decisions to our classification rule by considering {multiple decision neurons instead of a single neuron. The resulting readout is arguably the simplest combination of decision neurons, where the computation is distributed across the linear classifiers.} The model we study is also closely related to the recently proposed Gated Linear Networks \cite{budden2020gaussian,Sezener2021.03.10.434756,veness2021gated}: nonlinear neural architectures that remain analytically tractable \cite{saxe2022neural,li2022globally}. 
In particular, context switching is implemented by a gating mechanism, where the gating functions depend on a set of ``context vectors" $\rr_k\in\mathbb{R}^N$, ${k=1,\ldots K}$. The network output is
\begin{align}
f_{\bm{W}}(\xx)=\sum_{\cc\in M}g_{\cc}\left(\{\xx^\top\rr_k\}_{k=1}^K\right) \,{\ww_\cc^\top\xx}\;,\label{eq:equation}
\end{align}
where $\xx$ denotes the input, $\bm{W}=\{\ww_\cc\}_{\cc\in M}$ are the trainable weights, each associated to a context $\cc\in M$. The label estimate is given by $\hat y =\sign\left(f_{\bm W}(\xx)\right)$. The gating functions $g_\cc$ partition the input space into different regions---that can be interpreted as receptive fields or contexts---so that $g_\cc$ equals one if and only if $\xx$ belongs to context $\cc$, and zero otherwise. Notably, our theory holds for generic gating functions that partition the input space, ensuring that contexts do not overlap and each input is assigned to a single, distinct context. Furthermore, the number of contexts $K$ does not have equal to the number of decision neurons $|M|$. For concreteness, we will focus on \emph{half-space gating} \cite{veness2021gated}, where an exponential number $|M|=2^K$ of non-overlapping contexts $\cc\in M=\{0,1\}^K$ result from $K$ binary decisions implemented by the context hyperplanes: $g_\cc(\{\xx^\top\rr_k\}_{k=1}^K) =\prod_{k=1}^K\delta_{\Theta(\rr_k^\top\xx),c_k}$, where $\delta_{i,j}$ denotes the Kronecker $\delta-$function and $\Theta$ the Heaviside step function. Further details are provided in the Supplemental Material (SM) \cite{supmat} (including references \cite{gardner1988space,mezard1987spin,chung2018classification,wakhloo2023linear,aubin2018committee,loureiro2021learning,cornacchia2023learning,rosch1975family,sorscher2022neural,cover1965geometrical,chou2024,stephenson2021geometry,cohen2020separability}), where we also discuss another example of gating function. The context hyperplanes are fixed---possibly resulting from a previous optimization---while biases can be easily incorporated into the gating functions. Fig.~\ref{fig:contexts1} illustrates half-space gating in a non-linearly-separable classification task where representations of the same concept correspond to different sensory modalities or stylistic frameworks. The interplay of manifold and context geometry is schematized in Fig.~\ref{fig:illustration}. 
\subsection{Manifold capacity}
A key quantity to assess the efficiency of neural representations is the manifold capacity, i.e., the maximal number of manifolds per dimension $\alpha=P/N$ that can be correctly classified by the nonlinear rule in Eq.~\eqref{eq:equation} with high probability at a given margin $\gamma\geq 0$. 
In particular, we consider the thermodynamic limit $P,N\rightarrow\infty$, with $\alpha=\mathcal{O}_N(1)$. This corresponds to finding the largest $\alpha$ such that there exist a collection of decision hyperplanes $\bm{W}$, $\Vert \ww_\cc\Vert_2 ^2=N$ for all $\cc\in\{0,1\}^K$, satisfying: $\min_{\xx^\mu\in\mathcal{M}^\mu}y^\mu f_{\bm{W}}(\xx^\mu)\geq \gamma$ for all $\mu=1,\ldots,P$ with probability one in the thermodynamic limit.
This threshold can be determined by computing the average logarithm of the Gardner volume \cite{gardner1988space}---the volume of the space of solutions---in the thermodynamic limit:
\begin{align}
\begin{split}
    \overline{\ln V}=\\\hspace{-.4em}\overline{{\left(\prod_{\cc\in\{0,1\}^K}\hspace{-.5em}\underset{\mathcal{S}(\sqrt{N})}{\int} \dd\ww_\cc\right)}\hspace{-.2em}\prod_{\mu=1}^P\hspace{-.2em}\Theta\hspace{-.2em}\left(\min_{\xx^\mu\in\mathcal{M}^\mu}y^\mu f_{\bm{W}}(\xx^\mu)-\gamma\right)},
\end{split}
\end{align}
where $\mathcal{S}(\sqrt{N})$ is the $(N-1)-$dimensional hypersphere of radius $\sqrt{N}$ and the overbar denotes the average with respect to the labels $\yy$ and the manifold directions $\bm{U}$. We compute the Gardner volume using the replica method \cite{mezard1987spin,monasson1995learning,baldassi2019properties,zavatone2021activation}. We defer the details of this computation to the SM \cite{supmat}. We derive an exact formula for the storage capacity $\alpha^*$ for our model in the thermodynamic limit: 
\begin{align}
\begin{split}
\label{eq:formula_manifolds}
    \frac{1}{\alpha^*(K,\bm{\Phi},\gamma)}=\\\hspace{-.3em}\mathbb{E}_{{\bm{y},\bm{\xi},\RR}}\left[\max_{\cc\in\{0,1\}^{K}}\underset{{{\bm{H}}_\cc\in \mathcal{H}^\gamma_\cc(\yy,\bm{\Sigma}|\bm{R})}}{\min}\displaystyle   \;\frac 1P\sum_{\mu=1}^P{\Vert \bm{H}^\mu_{\cc}-\xxi^\mu_\cc\Vert_2^2}\right].
\end{split}
\end{align}
\begin{figure}[t!]
\includegraphics[scale=.32]{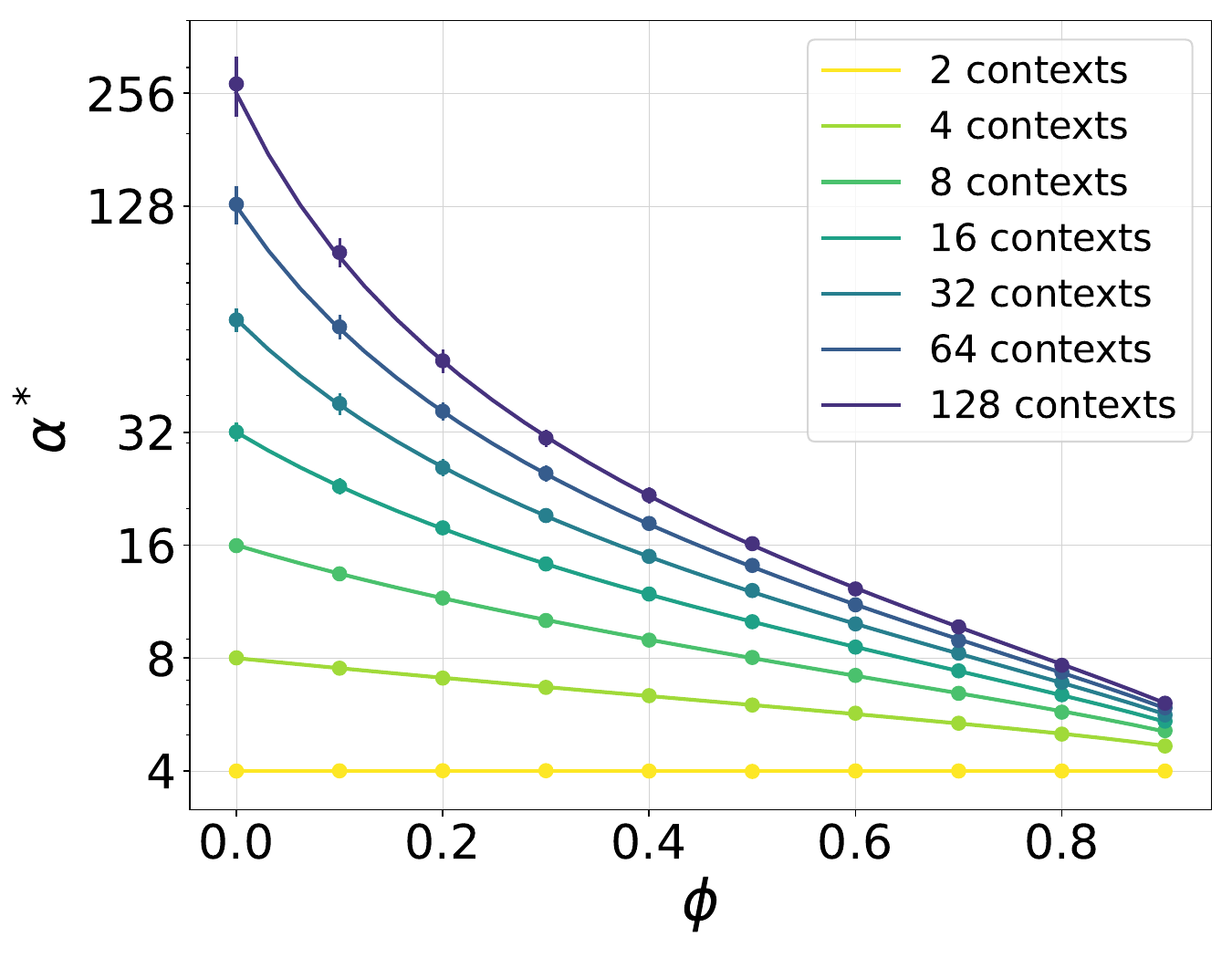}
    \caption{Capacity as a function of the context correlation parameter $\phi$ in the case of uncorrelated random points. Curves for different number of contexts are depicted with different colors. Full lines mark our theoretical predictions from Eq.~\eqref{eq:formula_randomPoints}, symbols mark numerical simulations at $N=5000$.}
    \label{fig:formula}
\end{figure}
The capacity depends on the context correlations and the manifold statistics via the local fields:
\begin{align}
    {H}^\mu_{i,\cc} \coloneqq y^\mu\frac{\ww_\cc^\top \uu^\mu_i}{\sqrt{N}} \;, && {R}^\mu_{i,k}\coloneqq\frac{\rr_k^\top\uu^\mu_i}{\sqrt{N}}\;, 
\end{align}
with $\mu=1,\ldots P$, $i=0,\ldots,D$, $k=1,\ldots, K$, and $\cc\in\{0,1\}^K$. In particular, ${H}^\mu_{i,\cc}$ represents
the local field induced by the solution vector $\ww_\cc$ on the basis vector $\uu^\mu_i$. Similarly, ${R}^\mu_{i,k}$ corresponds to the local field induced by the context-assignment vector $\rr_k$ on the basis vector $\uu^\mu_i$, and $\bm{R}^\mu_i\sim\mathcal{N}\left(\bm{0},\bm{\Phi}\right)$. The matrix $\bm{\Phi}\in\mathbb{R}^{K\times K}$ encodes the correlations between different context hyperplanes: 
$\Phi_{kk'}=\rr_k^\top\rr_{k'}/N$, with the normalization $\Vert \rr_k\Vert ^2=N$ for all $k=1,\ldots,K$. The local fields ${H}^\mu_{i,\cc}$ and ${R}^\nu_{j,k}$ are coupled through the optimization constraint $
\HH_\cc\in\mathcal{H}^\gamma_\cc(\yy,\bm{\Sigma}|\RR)$, which we discuss in further detail in the SM \cite{supmat}. Finally, the Gaussian variable ${\xi}_{i,\cc}^\mu\sim\mathcal{N}(0,1)$
encodes the part of the variability in ${H}^\mu_{i,\cc}$ due to quenched
variability in the basis vector $\uu^\mu_i$. 
\begin{figure*}
\hspace{-1.5em}\includegraphics[scale=0.3]{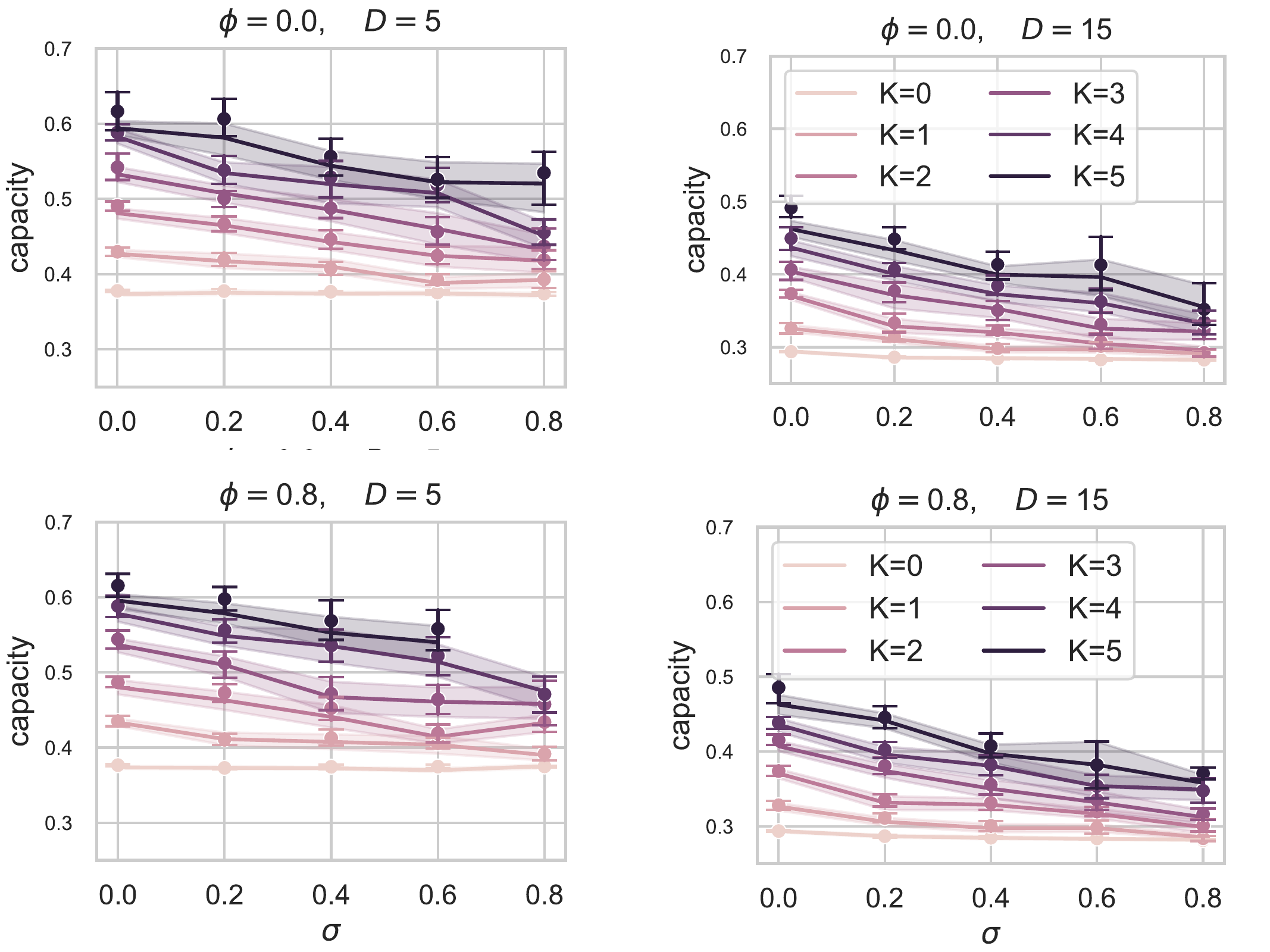}\caption{Capacity as a function of the manifold correlation, that for visibility purposes we take uniform $\Sigma^{\mu i}_{\nu j}=\sigma$ for all pairs of manifolds $(\mu,\nu)$ and directions $(i,j)$ for synthetic spherical manifolds. Subplot in different rows represent different values of context correlation $\phi\in [0,0.8]$, while different columns represent different latent dimension $D\in [5,15]$, embedded in ambient dimension $N=4000$. We consider $P=50$ spherical manifolds, and for each we draw $M=50$ points. Each panel depicts the capacity for $2^K=1,2,4,8,16,32$ contexts, represented by different colors. Full lines mark theoretical predictions while dots mark simulations.   }
    \label{fig:theoryVSexperiments}
\end{figure*}

In summary, context dependence introduces additional constraints into the optimization problem, coupling the manifold geometry to the structure of the context hyperplanes. We thoroughly test the validity of our theory on a synthetic dataset of spherical manifolds. Fig.~\ref{fig:theoryVSexperiments} shows our theoretical prediction for the capacity for different manifold correlations and number of contexts, while a broader set of parameters is shown in the SM \cite{supmat}. We find excellent agreement between theory and numerical simulations. 
\section{Examples on synthetic data} 
In this section, we analyze synthetic data to explore how the interaction between manifold structure and context correlations influences capacity. 
\subsection{Special case: random points} We start by considering the simplest scenario of random points, i.e., $D=0$ and no manifold correlations, which proves useful for developing intuition on the problem. 
In particular, we take the context correlation matrix $\bm{\Phi}=(1-\phi)\bm{I}_K + \phi \mathbf{1}_K\mathbf{1}_K^\top$, where $\mathbf{1}_K$ denotes the $K-$dimensional vector with all entries equal to one. This choice allows to control context correlations tuning just one parameter $\phi\in[0,1]$. In this special case, the capacity formula for half-space gating functions reduces to 
\begin{align}
\begin{split}
    \frac{1}{\alpha^*(K,\phi)}=\frac{(1+\gamma^2)}{2^{K+1}}\\\hspace{-.8em}\times\hspace{-.3em}\underset{\cc\in\{0,1\}^K}\max\mathbb{E}_{\eta}\left[\prod_{k=1}^K\left[1+(1-2c_k) \erf\left(\frac{\sqrt{\phi}\eta}{\sqrt{2(1-\phi)}}\right)\right]\right]\hspace{-.2em},\label{eq:formula_randomPoints}
    \end{split}
\end{align}
with $\eta\sim\mathcal{N}(0,1)$. The details of this computation can be found in the SM \cite{supmat}. In this prototypical setting, we find analytically that the capacity grows at most linearly with the number of contexts, and the maximal capacity $\alpha^*\vert_{\gamma=0}=2^{K+1}$ is achieved at $\phi=0$, i.e., orthogonal contexts, as a generalization of the classical result in the absence of contexts: $\alpha^*\vert_{\gamma=0}=2$ derived by Cover \cite{cover1965geometrical}. This finding holds in the absence of biases, and aligns with the experimental observation that orthogonal brain representations support efficient coding \cite{FLESCH20221258,nogueira2023geometry}. From an ``efficient coding" perspective, enhanced capacity at low context correlations may result from the optimal tiling of the input space with contexts when stimuli exhibit minimal structure. This phenomenon is related to the long-standing idea that the statistics of stimuli shape the distribution of receptive fields \cite{david2004natural} to achieve coding efficiency \cite{doi2012efficient}. However, while prior work on efficient coding has focused on minimizing the loss of information, here we reframe the problem in terms of the task efficiency and capacity of the readout. Fig.~\ref{fig:formula} depicts the theoretical predictions from Eq.~\eqref{eq:formula_randomPoints} (full lines) and the numerical estimates (symbols) of the capacity as a function of the context correlation $\phi$, for increasing number of contexts.  

\subsection{Spherical manifolds}
We extensively check the validity of the capacity formula \eqref{eq:formula_manifolds} in a synthetic setting where neural manifolds are generated from the model in Eq.~\eqref{eq:u_distrib}. Specifically, correlated synthetic manifolds are generated in three steps. First, we randomly and independently sample the center vector and axes vectors for each manifold. Next, we introduce center and internal axes correlations to the manifolds by correlating the entries of these vectors. Finally, we sample random points on the sphere (specified by the center and axes vectors) of each manifold. 

Fig.~\ref{fig:theoryVSexperiments} shows the comparison between theory and simulations for half-space gating functions and random context hyperplanes. Each panel depicts the capacity as a function of manifold correlation, varying the number of contexts. In order to describe manifold correlation via just one control parameter, we set all the off-diagonal entries of $\bm{\Sigma}$ to the same value $\sigma$, that we plot on the $x-$axis:  $\Sigma^{\mu i}_{\mu i}=1$, $\Sigma^{\mu i}_{\nu j}=\sigma$ for all $(\mu,i)\neq (\nu,j)$. The context correlation parameter is $\phi=0,0.8$ in the upper and lower panel respectively, while the latent dimension is $D=5,15$ from left to right. A broader range of parameters is shown in the SM \cite{supmat}. We find an excellent agreement between our theoretical predictions---marked by full lines--and numerical simulations---marked by dots. We provide more details on the numerics in the SM \cite{supmat}.

\begin{figure}[t!]
\includegraphics[width=.4\textwidth]{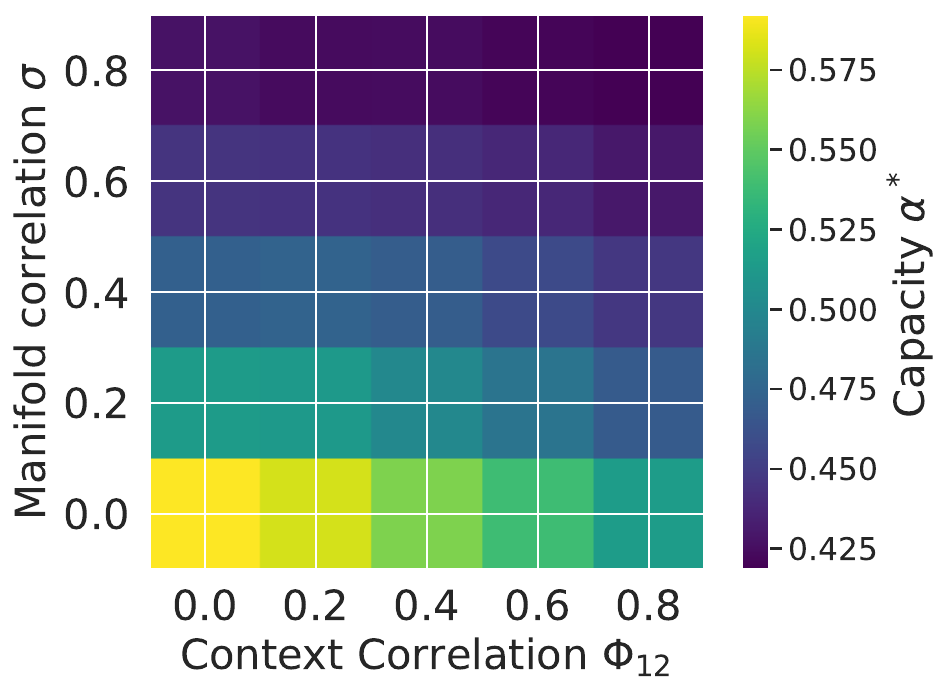}
    \caption{Capacity from Eq.~\eqref{eq:formula_manifolds} for spherical synthetic manifolds of latent dimension $D=10$, embedded in ambient dimension $N=8000$, different combinations of uniform manifold and context correlations $(\sigma,\Phi_{12})$, and four contexts.}
    \label{fig:heatmap}
\end{figure}

We further illustrate the interplay of manifold and context correlations in Fig.~\ref{fig:heatmap}. We consider four random contexts, i.e., $K=2$ context hyperplanes, and spherical synthetic manifolds of latent dimension $D=10$. We plot values of the capacity from Eq.~\eqref{eq:formula_manifolds} for a range of $\Phi_{12}$ and $\sigma$. We find that the capacity is decreasing both in manifold and context correlations. 
In particular, we find that increasing manifold correlations leads to a significant drop in capacity when random contexts are used.
\subsection{Gaussian mixture model}
\begin{figure}[t!]
    \centering
    \includegraphics[width=\linewidth]{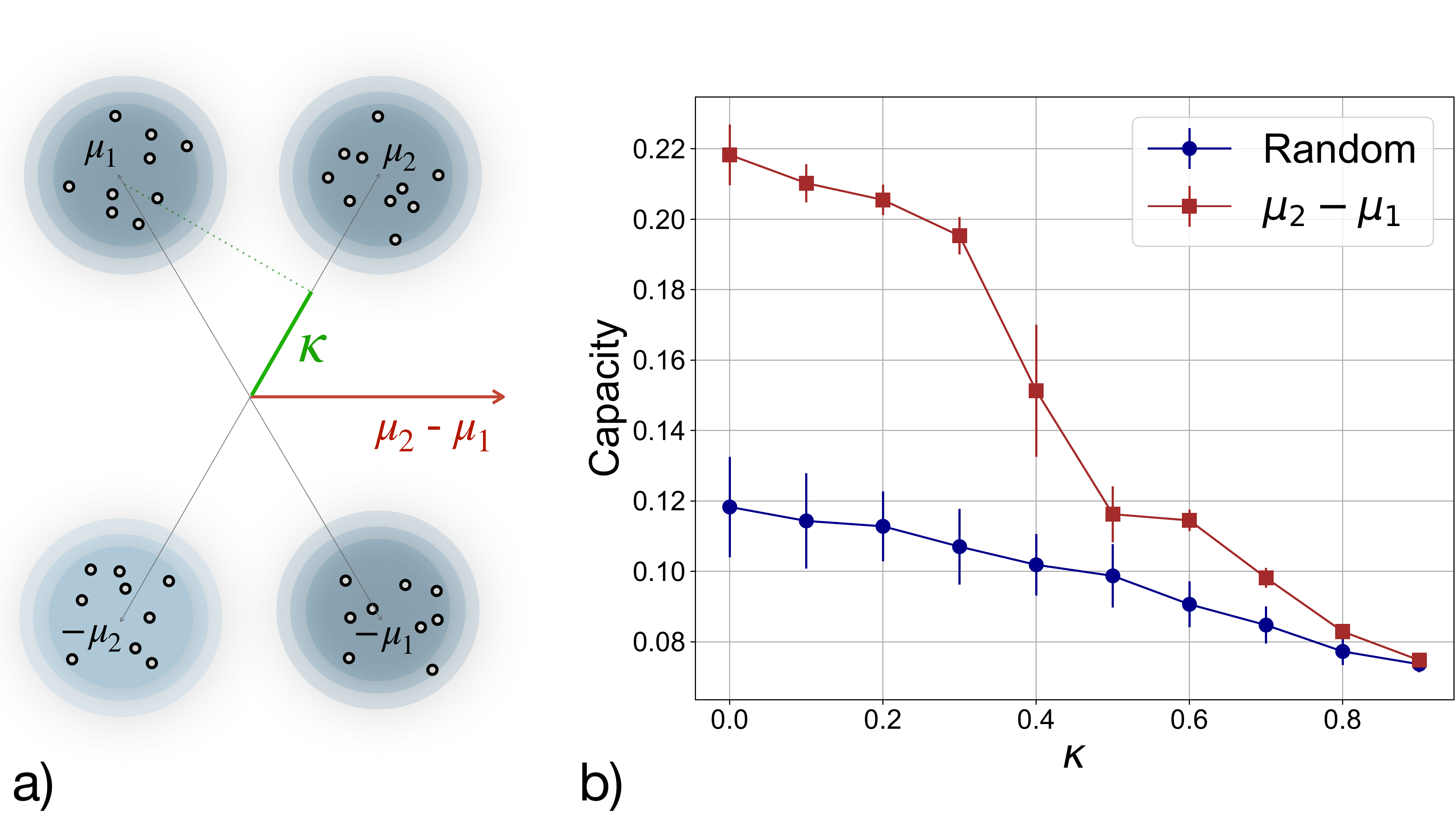}
    \caption{\textbf{a)} Each manifold is a point cloud drawn from a Gaussian centered at $\pm\bm{\mu}_{j}$, $j=1,2$. The alignment between the centers is marked by $\kappa$. The red arrow marks the context vector $\bm{r}_1=\bm{\mu}_2-\bm{\mu}_1$. \textbf{b)} Manifold capacity as a function of the alignment $\kappa$ for a random context vector (blue) and $\bm{r}_1$ (red). Each cloud has $40$ points, in dimension $N=200$. We plot average and standard deviation over $10$ realizations of the manifolds and random context.}
    \label{fig:xor}
\end{figure}

It would be interesting to investigate whether these trends change when context assignments are optimized to maximize capacity. Indeed, this procedure could serve as a powerful lens to uncover latent informative directions in high-dimensional neural representations. While we leave a detailed investigation of this direction for future work, we present a prototypical example to demonstrate that specific choices of context hyperplanes can enhance capacity compared to random ones. As an illustrative case, we take four Gaussian point clouds as neural manifolds. Each cloud is centered at one of the unit-norm vectors $\pm\bm{\mu}_i$ ($i=1,2$). The cosine similarity between $\bm{\mu}_1$ and $\bm{\mu}_2$ is controlled by a scalar parameter $\kappa$. Intuitively, increasing $\kappa$ reduces the separability of the point clouds. A pictorial representation of this setting is given in panel \textbf{a)} of Fig.~\ref{fig:xor}. We compute the context-dependent capacity with half-space gating and $K=1$ context vector. Panel \textbf{b)} compares the capacity as a function of $\kappa$ for a random context vector to that of a linear combination of the centers: $\bm{r}_1=\bm{\mu}_2-\bm{\mu}_1$. We average the capacity over $10$ realizations of the manifolds and the random context. We find that this natural choice of context hyperplane results in a capacity increase compared to the random case, indicating that leveraging ``relevant" directions in the data improves readout efficiency.

\begin{figure*}[t!]
    \centering
    \includegraphics[scale=.32]{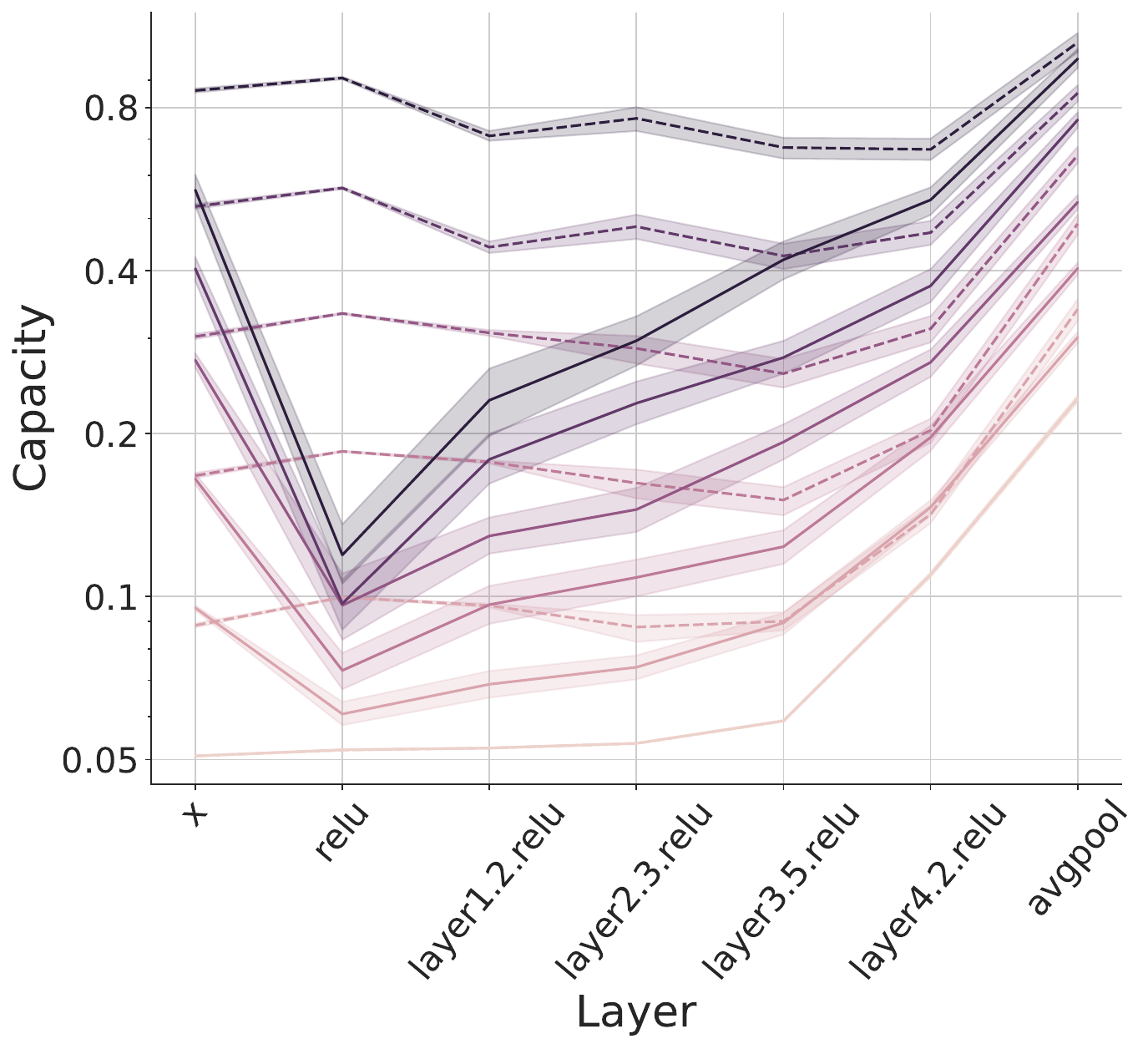}
    \includegraphics[scale=.32]{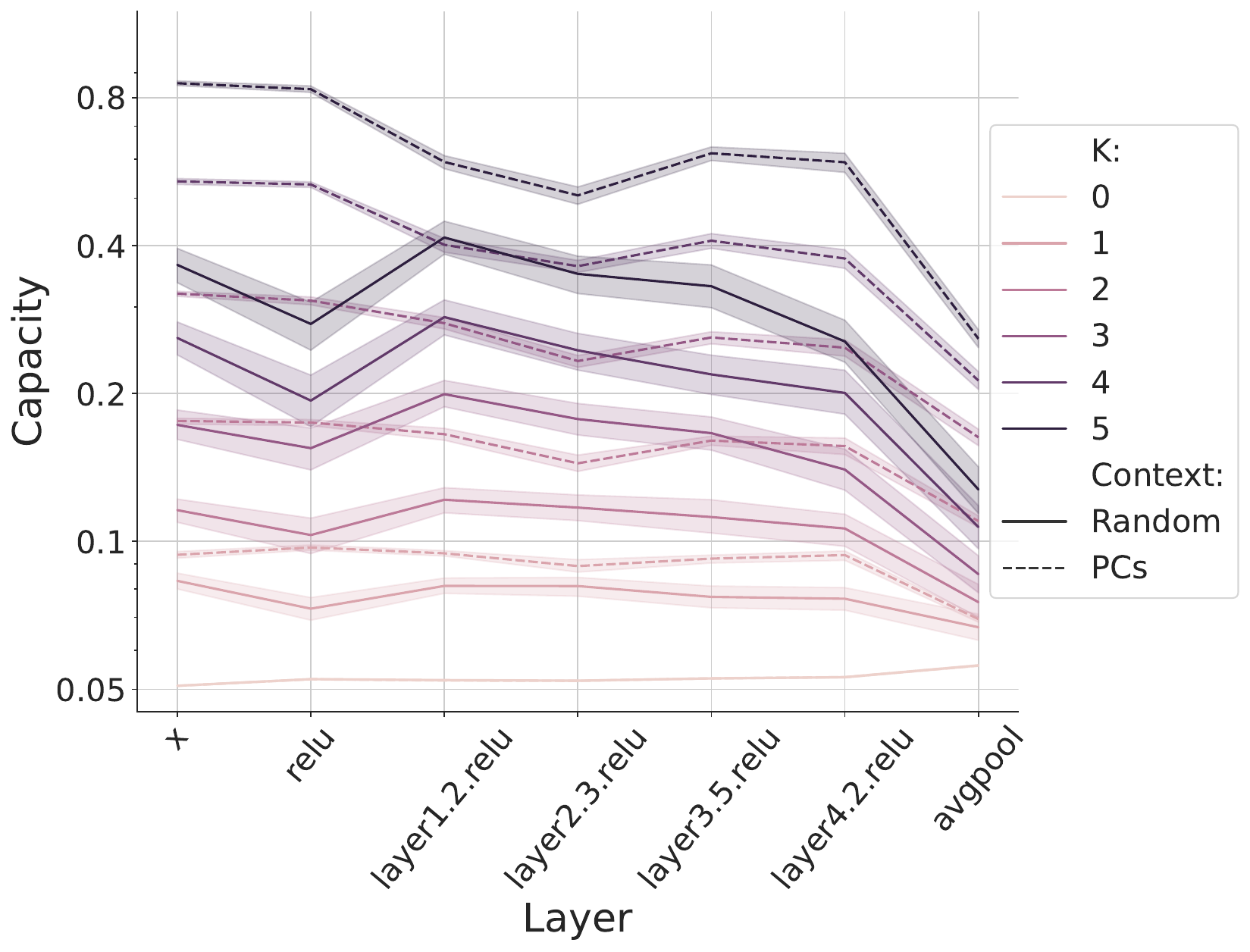}
    \caption{Capacity across layers of ResNet-50 pretrained on ImageNet, under a supervised objective (left panel) and a SimCLR objective (right panel). Different colors represent the number of contexts $2^K=1,2,4,8,16,32$. Solid lines indicate random contexts, while dashed lines represent principal components used as context vectors.  }
    \label{fig:applications}
\end{figure*}
\section{Applications on real data}
We next apply our theoretical framework to quantify the capacity of neural representations in deep neural networks. We carry out experiments on Resnet-50 \cite{he2016deep}
pretrained on ImageNet with supervised and SimCLR objectives. We consider 50 image classes, randomly selected, and draw 50 images for each class to form neural manifolds, following the procedure in~\cite{cohen2020separability}. We compute the capacity using the replica formula in Eq.~\eqref{eq:formula_manifolds}.

Fig.~\ref{fig:applications} displays the capacity as a function of the layer depth for different number of contexts $2^K=1,2,4,8,16,32$. The specific layers at which we capture the representations are indicated on the $x-$axis. 
The left panel shows results for the supervised objective, while the right panel shows results for SimCLR. 
\subsection{Random context vectors}
First, we consider random uncorrelated context hyperplanes ($\bm{\Phi}=\bm{I}_K$), marked by full lines. Even in the simplest case of random contexts, we find that our framework can quantify the progressive restructuring and disentanglement of representations across layers, while the capacity curve in the absence of contexts ($K=0$) remains relatively flat until the last layers. 
Interestingly, even with just two contexts $(K=1)$, we can detect early-layer representation restructuring, a phenomenon that becomes more pronounced as additional contexts are incorporated. This approach also uncovers distinct capacity trends for the two objectives. For the supervised objective, capacity exhibits an initial compression from the input to the first hidden layer, followed by a monotonic increase, reflecting progressive representation untangling. In contrast, the SimCLR objective shows a nonmonotonic pattern, with capacity rising in the earlier layers but decreasing toward the readout, suggesting a different strategy for encoding representations. None of these phenomena could be captured by the context-less linear capacity theory \cite{chung2021neural}.

The number of contexts required to achieve a given capacity approximates the number of piece-wise linear components of the decision boundary, hence it is an indirect measure of the complexity of the representation. If we focus on the supervised objective and set, e.g., $\alpha=0.1$ we find that the readout layer can achieve this separability threshold even at $K=0$, while this number increases going back to the representations in the first hidden layer that requires $2^K=16$ contexts to be shattered. This observation quantifies the level of representation disentanglement as a progressive linearization of the decision boundary.
\subsection{Principal components as context vectors}
As demonstrated with synthetic manifolds in the illustrative example of Fig.~\ref{fig:xor}, structured representations are often shaped by latent directions that encode meaningful information. Therefore, it is interesting to investigate how the capacity trends change if special directions in the data are chosen to define contexts. To this end, we repeat the analysis of ImageNet representations using the principal components (PCs) of the data as context vectors. Specifically, we exclude the first PC, which assigns all points to the same context, and instead use the $K$ PCs starting from the second. The capacity across layers is plotted with dashed lines in Fig.~\ref{fig:applications} for supervised objective (left panel) and SimCLR objective (right panel). Crucially, we find that, in all cases, the capacity computed using PCs to gate representations is larger than the one computed with random contexts. Furthermore, the PC-capacity trends are significantly different than those obtained for random contexts. 

These findings support the intuition that informative directions in the data serve as meaningful axes to enhance representation efficiency. Conversely, it suggests a novel approach for identifying these special latent directions as those maximizing the capacity of the representation.
\section{Discussion}
We have adopted
context dependence as a unifying lens to study the efficiency of neural representations for complex tasks. 
In particular, we have proposed an analytically-solvable model 
that incorporates contextual information into classification, allowing for nonlinear decision boundaries. The decision-making rule is {distributed across a} collection of input-dependent ``expert" neurons, each associated with distinct contexts via gating mechanisms.   
 First, we have derived an analytic expression for the manifold capacity. This formula elucidates the interplay between stimulus statistics and correlations within context hyperplanes. Furthermore, it enables probing the manifold capacity for non-linear readouts via piece-wise linear decision boundaries.  Applications to both synthetic examples and artificial neural networks demonstrate the validity of our theoretical predictions. Our theory allows to explore the structure of memorization across layers of deep neural networks, capturing nonlinear processing of representations. 
 \\\indent The framework presented here paves the way for various further investigations. From a theoretical perspective, a natural extension of this work involves developing learning algorithms able to find high-capacity context assignments. {A first step in this direction is the study of gradient-descent training of contexts and decision-making neurons, with possibly different objectives, to balance memorization and generalization.} A complementary perspective involves exploring the interplay between context and manifold geometry by explicitly introducing correlations between context vectors and manifold directions in the model. In particular, it would be interesting to explore how contextual information impacts the classification of hierarchical data structures. Finally, it would be interesting to explore how different gating functions impact the capacity. An alternative to half-space gating is discussed in the SM \cite{supmat}.
 
 On the applications side, we plan to leverage our theory to test the efficiency of neural representations from biological datasets. To this end, it would be relevant to extend the analysis of \cite{chou2024} by deriving effective geometric measures for manifold capacity with contexts. This extension would allow a systematic application of our theory to large-scale neural recordings. 
While in this paper we focus on contextual information gated from the input stimuli side, there have been lines of works studying contextual readout, e.g., multi-tasking in cognitive control~\cite{musslick2017multitasking}. Extending our theory to unify the role of contextual information in both input and output can enhance our understanding of how neural representations efficiently accommodate complicated tasks.
\vspace{1em}
\subsection*{Acknowledgments} We thank Will Slatton for useful feedback on this manuscript. This work was funded by the Center for Computational Neuroscience at the Flatiron Institute of the Simons Foundation. S.C. is supported by the Klingenstein-Simons Award, a Sloan Research Fellowship, NIH award R01DA059220, and the Samsung Advanced Institute of Technology (under the project "Next Generation Deep Learning: From Pattern Recognition to AI"). All experiments were performed on the Flatiron Institute high-performance computing cluster. F.M. was supported by the Simons Foundation (Award Number: 1141576).

\bibliographystyle{apsrev4-1} 
\bibliography{biblio} 
\appendix
\newpage
\onecolumngrid

\renewcommand{\theequation}{S\arabic{equation}}
\renewcommand{\thetable}{S\arabic{table}}
\renewcommand{\thefigure}{S\arabic{figure}}
\setcounter{equation}{0}
\setcounter{table}{0}
\setcounter{figure}{0}

\section{Derivation of the Gardner volume via the replica method}
\label{appendix:replica}
In this section, we provide additional details on the derivation of the capacity threshold using the replica method \cite{gardner1988space,mezard1987spin}. We use the replica trick to compute the log-volume averaged over the manifold directions $\bm{U}=\{\bm{u}_i^\mu\}_{i=0,\ldots,D}^{\mu=1,\ldots,P}$ and the labels $\yy=\{y^\mu\}_{\mu=1}^P$. We denote this average by an overbar: $\overline{\ln Z}=\lim_{n\rightarrow 0^+}\left(\overline{Z^n}-1\right)/n$. The averaged $n^{\rm th}$ moment of the partition function is
\begin{align}
\begin{split}
    \overline{Z^n}=\mathbb{E}_{\UU,\yy}\left[\int \left(\prod_{a=1}^n \prod_{\cc\in\{0,1\}^K}\dd \ww^a_\cc \;\delta\left(N-\Vert \ww^a_m\Vert_2^2\right)\right)\left[\prod_{\mu=1}^P\prod_{\cc\in\{0,1\}^{K}}\Theta\left(J^{(\cc)}_{\mathcal{S}}(\HH^{\mu,a},\RR^\mu)-\gamma\right)\right]\right]\;,
    \end{split}
    \end{align}
where we have used the definitions
\begin{align}
    H^{\mu,a}_{i,\cc}\coloneqq y^\mu\frac{{\ww^a_\cc}^\top \uu_i^\mu}{\sqrt{N}}\;,&&
R^{\mu}_{i,k}\coloneqq\frac{{\rr_k}^\top\uu^\mu_i}{\sqrt{N}}\;,\label{eq:auxiliaryHR}
\end{align}
while the function $J^{(\cc)}_{\mathcal{S}}$ encodes the optimization constraints
\begin{align}
    J^{(\cc)}_{\mathcal{S}}(\HH,\RR)\coloneqq\begin{cases}
     H_{0,\cc}+\underset{\bm{s}\in \mathcal{S}\cap \mathcal{C}_{\cc|\RR}}{\min}\left\{ {\HH_\cc}^\top\ss\right\}& \text{if } \mathcal{S}\cap \mathcal{C}_{\cc|\RR}\neq\emptyset \\ \bar J >\gamma& \text{if } \mathcal{S}\cap \mathcal{C}_{\cc|\RR}=\emptyset 
    \end{cases}\;.
\end{align}
For each context $\cc$, the manifold coordinates $\ss$ are constrained to lie on an effective context, here denoted by $\mathcal{C}_{\cc|\RR}=\{\ss \in\mathbb{R}^{D+1}:g_\cc\left(\{\RR_k\}_{k=1}^K\right)=1\}$. While in the main text we focus on the half-space and prototype gating functions, here we derive the capacity threshold for generic gating functions, as long as they define non-overlapping contexts. The constant $\bar{J}$ is introduced to ensure that the Heaviside-$\Theta$ function is $1$ when $\mathcal{S}\cap \mathcal{C}_{\cc|\RR}=\emptyset$.
We can introduce the auxiliary variables defined in Eq.~\eqref{eq:auxiliaryHR} via the Fourier transform of the Dirac $\delta-$function:
    \begin{align}
  \begin{split}
   \overline{Z^n} =\int \left(\prod_{a=1}^n \prod_{\cc\in\{0,1\}^K}\dd \ww^a_\cc \;\delta\left(N-\Vert \ww^a_\cc\Vert_2^2\right)\right)\\\times \mathbb{E}_{\bm{U},\yy}\left[\prod_{\mu=1}^P\int \mathcal{D}\RR^{\mu}\prod_{a=1}^n\mathcal{D}\HH^{\mu,a}\prod_{i=0}^{D}\left[\prod_{\cc\in\{0,1\}^K}\sqrt{2\pi}\;\delta\left(H^{\mu,a}_{i,\cc}-y^\mu\frac{{\ww^a_\cc}^\top \uu_i^\mu}{\sqrt{N}}\right)\right]\left[\prod_{k=1}^{K}\sqrt{2\pi}\;\delta\left(R^\mu_{i,k}-\frac{\rr_k^\top\uu^\mu_i}{\sqrt{N}}\right)\right]\right],
\end{split}
\end{align}
and 
\begin{align}
\mathcal{D}\RR^{\mu}\prod_{a=1}^n\mathcal{D}\HH^{\mu,a}\coloneqq\left[\prod_{i=0}^{D}\left(\prod_{k=1}^{K}\frac{\dd R^\mu_{i,k}}{\sqrt{2\pi}}\right)\left(\prod_{a=1}^n\prod_{\cc\in\{0,1\}^K}\frac{\dd H^{\mu,a}_{i,\cc}}{\sqrt{2\pi}}\right)\right]\prod_{a=1}^n\prod_{\cc\in\{0,1\}^{K}}\Theta\left(J^{(\cc)}_{\mathcal{S}}(\HH^{\mu,a},\RR^\mu)-\gamma\right)\;.
\end{align}
We assume that the vectors defining the manifold directions are drawn from the multivariate Gaussian distribution in Eq.~(1) of the main text. Hence the variables $\HH^{\mu,a}$ and $\RR^\mu$ are Gaussian, with zero mean and covariance
\begin{align}
    \mathbb{E}_\UU\left[ H^{\mu,a}_{i,\cc} H^{\nu,a'}_{j,\cc'}\right]=y^\mu y^\nu Q^{aa'}_{\cc \cc'}\Sigma^{\mu i}_{\nu j}\;,&&
    \mathbb{E}_\UU\left[  H^{\mu,a}_{i,\cc} R^{\nu}_{j,k}\right]=y^\mu \rho^{a}_{\cc,k}\Sigma^{\mu i}_{\nu j}\;,&&
    \mathbb{E}_\UU\left[  R^{\mu}_{i,k} R^{\nu}_{j,k'}\right]=\Phi_{kk'} \Sigma^{\mu i}_{\nu j}\;,
\end{align}
where we have defined the overlap parameters
\begin{align}
Q^{aa'}_{\cc\cc'}\coloneqq \frac{{\ww^a_\cc}^\top\ww_{\cc'}^{a'}}{N}\;,&& Q^{aa}_{\cc\cc}=1\;,&&
\rho^{a}_{\cc,k}\coloneqq\frac{{\ww^a_\cc}^\top \rr_k}{ N}\;,&&
\Phi_{kk'}\coloneqq\frac{\rr_k^\top\rr_{k'}}{{N}}\;.
\end{align}
We can introduce these definitions via Dirac $\delta-$functions and compute the integrals over the weights. The average partition function to the $n^{\rm th}$ power can be rewritten as
\begin{align}
    \begin{split}
        \overline{ Z^n }\propto\int \dd \bm{Q}\,\dd \bm{\rho}\,\ee^{\frac{N}{2}\ln\det\left(\bm{Q}-\sum_{k,k'}\Phi_{kk'}^{-1}\bm{\rho}_k\bm{\rho}_{k'}^\top\right)+\ln\mathcal{Z}}=\int \dd \bm{Q}\,\dd \bm{\rho}\,\ee^{S(\bm{Q},\bm{\rho})}\;,\label{eq:S_action_def}
    \end{split}
\end{align}
where
\begin{align}
\label{eq:actionS}
\mathcal{Z}=\int \left[\prod_{\mu=1}^P\mathcal{D}\bm{R}^\mu\prod_{a=1}^n\mathcal{D}\bm{H}^{\mu,a}\right]\exp\left(-\frac{1}{2} \sum_{\mu,\nu,i,j} {\mathbb{H}_{\mu i}}^\top (\mathbb{C}^{-1}\otimes \mathbb{Q}^{-1})^{\mu  i}_{\nu j}\,\mathbb{H}_{\nu j}-\frac{1}{2}\ln\det\left(\mathbb{C}\otimes\mathbb{Q}\right)\right)\;,
\end{align}
and we have introduced the shorthand notation:
\begin{align}
\mathbb{H}^{a,\cc,k}_{\mu i}=\left(\begin{array}{c}
         H^{\mu,a}_{i,\cc}  \\
          R^\mu_{i,k}
    \end{array}\right)\;,&& \mathbb{C}^{\mu i}_{\nu j}=\left(\begin{array}{cccc}
        y^\mu y^\nu \Sigma^{\mu i}_{\nu j} &  y^\mu \Sigma^{\mu i}_{\nu j} \\
     y^\nu \Sigma^{\mu i}_{\nu j}     &  \Sigma^{\mu i}_{\nu j} 
    \end{array}\right)\;,&& \mathbb{Q}^{a\cc k}_{a'\cc ' k'}=\left(\begin{array}{cccc}
       Q^{aa'}_{\cc\cc'} &   \rho^{a'}_{\cc',k'}\\
    \rho^{a}_{\cc,k}     & \Phi_{kk'}
    \end{array}\right)\;.
\end{align}
\paragraph{Replica symmetric ansatz:} We assume the following replica symmetric (RS) ansatz \cite{mezard1987spin}
\begin{align}
Q^{aa}_{mm'}&=Q^*_{mm'}\;, \qquad\forall a=1,\ldots,n\;, \qquad\forall m,m'=1,\ldots M\;,\\
Q^{aa'}_{mm'}&=q_{mm'}\;, \qquad \forall a\neq a'\;,a,a'=1,\ldots,n\;,\qquad \forall m,m'=1,\ldots M\;,\\
\rho^a_{m,k}&=\rho_{m,k}\;,\qquad \forall a=1,\ldots n\;.
\end{align}
with the additional normalization constraint: $Q^*_{mm}=1$. The RS assumption is motivated by the observation that within each context the solution space is convex, and contexts do not overlap. Notice that we do not assume symmetry between different contexts. We compute all the terms in the action $S(\bm{Q}, \bm{\rho})$ in Eq.~\eqref{eq:S_action_def} under the RS ansatz.
\begin{align}
\begin{split}
&\frac{1}{2}\ln\det\left(\bm{Q}-\sum_{k,k'}\Phi^{-1}_{kk'}\bm{\rho}_k\bm{\rho}_{k'}^\top\right)=\frac{1}{2}\left[(n-1)\ln\det\left(\bm{Q}^*-\bm{q}\right)+\ln\det\left(\bm{Q}^*-\bm{q}+n\left(\bm{q}-\bm{\rho}^\top\Phi^{-1}\bm{\rho}\right)\right) \right].
\end{split}
\end{align}
 The matrix $\mathbb{Q}$ and its inverse $\mathbb{Q}^{-1}$ have the same block structure 
\begin{align}
\mathbb{Q}=\left(\begin{array}{ccccc}
  \bm{Q}^* & \bm{q} &\ldots & \bm{q} &\bm{\rho}^\top \\
\bm{q}&\ddots  & \bm{q} &\bm{q} &\bm{\rho}^\top\\
\bm{q}&\ldots&\bm{q}&\bm{Q}^*&\bm{\rho}^\top\\
\bm{\rho}&\bm{\rho}&\bm{\rho}&\bm{\rho}&\bm{\Phi}
\end{array}\right)\;,\qquad \mathbb{Q}^{-1}=\left(\begin{array}{ccccc}
  \tilde{\bm{Q}}^* & \tilde{\bm{q}} &\ldots & \tilde{\bm{q}} &\tilde{\bm{\rho}}^\top\\
\tilde{\bm{q}}&\ddots  & \tilde{\bm{q}} &\tilde{\bm{q}} &\tilde{\bm{\rho}}^\top\\
\tilde{\bm{q}}&\ldots&\tilde{\bm{q}}&\tilde{\bm{Q}}^*&\tilde{\bm{\rho}}^\top\\
\tilde{\bm{\rho}}&\tilde{\bm{\rho}}&\tilde{\bm{\rho}}&\tilde{\bm{\rho}}&\tilde{\bm{\Phi}}
\end{array}\right)\;,
\end{align}
and the inverse elements can be computed from the relation $\mathbb{Q}\mathbb{Q}^{-1}={\rm Id}$. We find the following relations
\begin{align}
\begin{split}
\tilde{\bm{Q}}^*&=\left(\bm{Q}^*-\bm{q}\right)^{-1}-\left(\bm{Q}^*+(n-1)\bm{q}\right)^{-1}\bm{q}\left(\bm{Q}^*-\bm{q}\right)^{-1}\\&\quad+\left(\bm{Q}^*+(n-1)\bm{q}\right)^{-1}\bm{\rho}^\top\left(\bm{\Phi}-n\bm{\rho}\left(\bm{Q}^*+(n-1)\bm{q}\right)^{-1}\bm{\rho}^\top\right)^{-1}\bm{\rho}\left(\bm{Q}^*+(n-1)\bm{q}\right)^{-1}\;,
\end{split}
\\\label{eq:tilde_q}
\tilde{\bm{q}}&=\tilde{\bm{Q}}^*-\left(\bm{Q}^*-\bm{q}\right)^{-1}\;,\\
\tilde{\bm{\rho}}&=-\left(\bm{\Phi}-n\bm{\rho}\left(\bm{Q}^*+(n-1)\bm{q}\right)^{-1}\bm{\rho}^\top\right)^{-1}\bm{\rho}\left(\bm{Q}^*+(n-1)\bm{q}\right)^{-1}\;,\\
\tilde{\bm{\Phi}}&=\left(\bm{\Phi}-n\bm{\rho}\left(\bm{Q}^*+(n-1)\bm{q}\right)^{-1}\bm{\rho}^\top\right)^{-1}\;.
\end{align}
 We then consider the Cholesky decomposition of the correlation matrix $\bm{\Sigma}$, that satisfies $\sum_{\mu',i'}L^{\mu i}_{\mu' i'}L^{\mu' i'}_{\nu j}=\Sigma^{\mu i}_{\nu j}$. It is useful to perform the rotations $H^{\mu,a}_{i,\cc}\rightarrow y^\mu \sum_{\mu',i'}L^{\mu i}_{\mu' i'}H^{\mu', a}_{i',\cc}$ and $R^{\mu}_{i,k}\rightarrow  \sum_{\mu',i'}L^{\mu i}_{\mu' i'}R^{\mu'}_{i',k}$. This transformation allows us to decouple the indices $(\mu,i)$ in the Gaussian weight, coupling them in the constraint functions $J^{(\cc)}_\mathcal{S}$. We find
\begin{align}
\mathcal{Z}=\int\left[\prod_{\mu=1}^P\mathcal{D}\bm{R}^\mu\prod_{a=1}^n\mathcal{D}\bm{H}^{\mu,a}\right]\exp\left(-\frac{1}{2} \sum_{\mu,i} {\mathbb{H}_{\mu i}}^\top  \mathbb{Q}^{-1}\mathbb{H}_{\mu i}-\frac{\alpha N}{2}\ln\det\mathbb{Q}\right)\;,\end{align}
where we have now changed the definition
\begin{align}
    \mathcal{D}\bm{R}^\mu\prod_{a=1}^n\mathcal{D}\bm{H}^{\mu,a}\coloneqq\left[\prod_{i=0}^{D}\left(\prod_{k=1}^{K}\frac{\dd R^\mu_{i,k}}{\sqrt{2\pi}}\right)\left(\prod_{a=1}^n\prod_{\cc\in\{0,1\}^K}\frac{\dd H^{\mu,a}_{i,\cc}}{\sqrt{2\pi}}\right)\right]\prod_{a=1}^n\prod_{\cc\in\{0,1\}^{K}}\Theta\left(J^{(\cc)}_{\mathcal{S}}(y^\mu\LL^\mu\HH^{a},\LL^\mu\RR)-\gamma\right)\;,
\end{align}
and we have used the shorthand notation $\LL^\mu\RR\in\mathbb{R}^{K\times(D+1)}$ to indicate the matrix whose $(k,i)$ entry is $\LL^{\mu i}\RR_k=\sum_{\mu',i'}L^{\mu i}_{\mu' i'}\RR^{\mu'}_{k,i'}$. 
Upon exchanging the limits $n\rightarrow 0^+$ and $N\rightarrow \infty$, we find that the high dimensional limit of the log-volume density can be computed via saddle point method:
\begin{align}
    \begin{split}\frac{\overline{\ln Z}}N\underset{N\gg 1}{\;\longrightarrow\;}
    \underset{\bm{Q^*},\bm{q},\bm{\rho}}{\text{extr}}\;S_{\rm RS}(\bm{Q^*},\bm{q},\bm{\rho})\;,
\end{split}
\end{align}
where
\begin{align}
   S_{\rm RS}(\bm{Q^*},\bm{q},\bm{\rho})& =\frac 12 \left[\ln\det(\bm{Q^*}-\bm{q})+{\rm Tr}\left((\bm{Q^*}-\bm{q})^{-1}\left(\bm{q}-\bm{\rho}^\top\bm{\Phi}^{-1}\bm{\rho}\right)\right) \right]\\&\quad+\alpha\mathbb{E}_{\yy,\bm{\xi},\bm{R}}\left[\frac 1P\ln\mathbb{E}_{\bm{H}}\left[\prod_{\mu=1}^P\prod_{\cc\in\{0,1\}^{K}}\Theta(J^{(\cc)}_{\mathcal{S}}(y^\mu\LL^\mu\HH^{a},\LL^\mu\RR)-\gamma)\right]\right],
\end{align}
and the averages are over $\bm{\xi}^\mu_i\sim\mathcal{N}(0,I_M)$, $\bm{R}^\mu_i~\sim~\mathcal{N}\left(\bm{\rho}\bm{q}^{-1/2}\bm{\xi}^\mu_i,\bm{\Phi}-\bm{\rho}^\top\bm{q}^{-1}\bm{\rho}^\top\right)$, and  $\bm{H}^\mu_i\sim\mathcal{N}\left(\bm{q}^{1/2}\bm{\xi}^\mu_i,\bm{\bm{Q}^*-\bm{q}}\right)$. 

The main additional technical difficulty in our analysis, compared to previous related works \cite{chung2018classification,wakhloo2023linear}, is that for more than one context the asymptotic closed-form formulas at a generic $\alpha$ are provided in terms of a set of coupled self-consistent saddle-point equations on $2^K \times  2^K$ and $K \times 2^K$ dimensional matrix  order parameters, instead of just one scalar order parameter. This poses some challenges on the numerical evaluation of the solution, as previously pointed out in the context of learning problems \cite{aubin2018committee,loureiro2021learning,cornacchia2023learning}.  We overcome these difficulties leveraging the simplifications of the formulas when evaluated at capacity $\alpha=\alpha^*$ and exploiting the structure of the solution space to make useful assumptions on the order parameters. 

\paragraph{The structure of the solution space with non-overlapping contexts.}
Given that contexts are non overlapping, each solution $\{\ww_\cc\}_{\cc\in\{0,1\}^K}\in \mathcal{W}$ is the union of solutions across contexts $\mathcal{W}=\mathcal{W}_1\times\mathcal{W}_2\times\ldots \times\mathcal{W}_{2^K}$. This implies $Q^*_{\cc\cc'}=q_{\cc\cc'}$ for all $\cc\neq \cc'$ and at every $\alpha$. Therefore, when contexts do not overlap, the covariance $(\bm{Q}^*-\bm{q})$ of the $\HH-$fields is diagonal, with entries  $(1-q_{\cc\cc})$. Moreover, given that labels are random and contexts are disjoint, it is reasonable to assume $q_{\cc\cc'}=0$ for all $\cc\neq\cc'$ and $\bm{\rho}=\bm{0}$.
These observations imply that the $\HH-$fields are effectively uncorrelated across contexts, hence we can factorize the expectation and write
\begin{align}
    \begin{split}
\ln\mathbb{E}_{\bm{H}}\left[\prod_{\mu=1}^P\prod_{\cc\in\{0,1\}^{K}}\Theta(J^{(\cc)}_\mathcal{S}(y^\mu \LL^\mu\bm{H}_\cc,\RR)-\gamma)\right]=\sum_{\cc\in\{0,1\}^{K}}\ln\mathbb{E}_{\HH_\cc} \left[\prod_{\mu=1}^P\Theta(J^{(\cc)}_\mathcal{S}(y^\mu\LL^\mu\bm{H}_\cc,\LL^\mu\RR)-\gamma)\right]\;.
    \end{split}
\end{align}
We obtain
\begin{align}
\begin{split}
    S_{RS}(\{{q}_{\cc\cc}\})&=\frac 12 \sum_{\cc\in\{0,1\}^K}\left[\ln\det(1-q_{\cc\cc})+\frac{{q}_{\cc\cc}}{1-q_{\cc\cc}}\right]\\&\qquad+\alpha\mathbb{E}_{\bm{y},\xxi,\RR}\left[\frac 1P\sum_{\cc\in\{0,1\}^{K}}\ln\mathbb{E}_{\HH_\cc} \left[\prod_{\mu=1}^P\Theta(J^{(\cc)}_\mathcal{S}(y^\mu\LL^\mu\bm{H}_\cc,\LL^\mu\RR)-\gamma)\right]\right]\;,\label{eq:SRSsimplified}
\end{split}
\end{align}
where averages are now over $\bm{\xi}^\mu_i\sim\mathcal{N}(0,I_M)$, $\bm{R}^\mu_i~\sim~\mathcal{N}\left(\bm{0},\bm{\Phi}\right)$, and  ${H}^\mu_{i,\cc}\sim\mathcal{N}\left(\sqrt{{q}_{\cc\cc}}\,\xi^\mu_{i,\cc}\,,\,{1-{q}_{\cc\cc}}\right)$.

\subsection{General case: capacity for manifolds}
We introduce an auxiliary function
\begin{align}
\hat S_\text{min}=\frac 12\ln(1-q_{\tilde\cc\tilde\cc})+\frac 12\frac{q_{\tilde\cc\tilde\cc}}{1-q_{\tilde\cc\tilde\cc}}+\frac {\alpha}P\mathbb{E}_{\bm{y},\xxi,\RR}\left[\ln\mathbb{E}_{\HH_{\tilde\cc}} \left[\prod_{\mu=1}^P\Theta(J^{(\tilde\cc)}_\mathcal{S}(y^\mu\LL^\mu\bm{H}_{\tilde\cc},\LL^\mu\RR)-\gamma)\right]\right],\label{eq:Smin_hat}
\end{align}
where $\tilde\cc=\text{arg}\underset{\cc\in\{0,1\}}{\text{min}}\frac 1P\ln\mathbb{E}_{\HH_\cc} \left[\prod_{\mu=1}^P\Theta(J^{(\cc)}_\mathcal{S}(y^\mu\LL^\mu\bm{H}_\cc,\LL^\mu\RR)-\gamma)\right]$. The function $\hat S_{\rm min}$ is a lower bound on the replica symmetric action $S_{\rm RS}$. Indeed, for each context $\tilde\cc$:
\begin{align}
    \ln(1-q_{\tilde\cc\tilde\cc})+\frac{q_{\tilde\cc\tilde\cc}}{1-q_{\tilde\cc\tilde\cc}}\geq 0\;, &&  \ln(1-q_{\tilde\cc\tilde\cc})+\frac{q_{\tilde\cc\tilde\cc}}{1-q_{\tilde\cc\tilde\cc}}\leq \sum_{\cc\in\{0,1\}^K}  \left[\ln(1-q_{\cc\cc})+\frac{q_{\cc\cc}}{1-q_{\cc\cc}}\right]\;.
\end{align}
We also have that
\begin{align}
\begin{split}
  2^K \lambda (\yy,\xxi,\RR)\leq \sum_{\cc'}\frac 1P\ln\mathbb{E}_{\HH_{\cc'}} \left[\prod_{\mu=1}^P\Theta(J^{(\cc')}_\mathcal{S}(y^\mu\LL^\mu\bm{H}_{\cc'},\LL^\mu\RR)-\gamma)\right]\;,
  \end{split}
\end{align}
where we have defined 
\begin{align}
    \lambda (\yy,\xxi,\RR) &\coloneqq  \underset{\cc\in\{0,1\}}{\text{min}}\frac 1P\ln\mathbb{E}_{\HH_{\cc}} \left[\prod_{\mu=1}^P\Theta(J^{(\cc)}_\mathcal{S}(y^\mu\LL^\mu\bm{H}_\cc,\LL^\mu\RR)-\gamma)\right]\leq 0\;.
\end{align}
Furthermore, due to the spherical constraint on the weight space, the log-volume density is bounded in the infinite-dimensional limit. Therefore, there exists a constant $\bar s$ such that 
\begin{align}
\hat S_{\rm min}+2^{K-1}\lambda (\yy,\xxi,\RR)\leq S_{\rm RS}\leq \hat S_{\rm min} +2^{K-1}\bar s,
\end{align}
and the phase transition at the capacity threshold $S_{\rm RS}\rightarrow -\infty$ is controlled by the divergence of $\hat S_{\rm min}$ to $-\infty$. From now on, the calculation follows the lines of \cite{chung2018classification,wakhloo2023linear}. As $q_{\tilde\cc\tilde\cc}\rightarrow 1$ in Eq.~\eqref{eq:Smin_hat}, at leading order we have 
\begin{align}
    \ln(1-q_{\tilde\cc\tilde\cc})+ \frac{{q}_{\tilde\cc\tilde\cc}}{1-q_{\tilde\cc\tilde\cc}}\;\underset{q_{\tilde\cc\tilde\cc}\rightarrow 1}\simeq\;  \frac{1}{1-q_{\tilde\cc\tilde\cc}}\;,
\end{align}
and
\begin{align}
    \frac 1 P\ln\mathbb{E}_{\bm{H}_{\tilde\cc}}\left[\prod_{\mu=1}^P\Theta(J^{(\tilde\cc)}_\mathcal{S}(y^\mu \LL^\mu\bm{H_{\tilde\cc}},\LL^\mu\RR)-\gamma)\right]\simeq -\frac 12 \;\;\frac{\underset{{\bm{H}}_{\tilde\cc}\in  \mathcal{H}_{\tilde\cc}^\gamma(\yy,\bm{\Sigma}|\RR)}{\min}\displaystyle \frac 1 P\sum_{\mu=1}^P\Vert \bm{H}^\mu_{\tilde\cc}-\bm{q}_{\tilde\cc}^{1/2}\xxi^\mu\Vert_2^2}{1-q_{\tilde\cc\tilde\cc}}\;,\label{eq:approx_S2}
\end{align}
where we have defined
\begin{align}
    \begin{split}
\mathcal{H}^\gamma_\cc(\yy,\bm{\Sigma}|\RR)=\Big\{\bm{H}_\cc \in \mathbb{R}^{ P\times (D+1)}\;: \;\forall \mu=1,\ldots,P\;,\quad\min_{\ss^\mu_\cc\in\mathcal{S}\cap \mathcal{C}_{\cc|\LL^\mu\RR}} y^\mu\sum_{\mu',i'}L^{\mu i}_{\mu' i'}H^{\mu'}_{i',\cc}s^\mu_i\geq \gamma\Big\}\;,
    \end{split}
\end{align}
and we have used the shorthand notation $\LL^\mu\RR\in\mathbb{R}^{K\times(D+1)}$ to indicate the matrix whose $(k,i)$ entry is $\LL^{\mu i}\RR_k=\sum_{\mu',i'}L^{\mu i}_{\mu' i'}\RR^{\mu'}_{k,i'}$. 
Therefore, close to the transition, we have that 
\begin{align}
    \hat S_{\rm min}\underset{q_{\tilde\cc\tilde\cc}\rightarrow 1}\simeq \frac 12\mathbb{E}_{\yy,\xxi,\RR}\left[\frac{1}{1-q_{\tilde \cc\tilde\cc}}\left(1- \alpha\underset{{\bm{H}}_{\tilde\cc}\in  \mathcal{H}(\yy,\bm{\Sigma}|\RR)}{\min}\displaystyle \frac 1 P\sum_{\mu=1}^P\Vert \bm{H}^\mu_{\tilde\cc}-\xxi^\mu\Vert_2^2\right)\right].
\end{align}
The transition happens when the right hand side changes sign, which determines the capacity threshold
\begin{align}
\label{eq:capacity_supmat}
    \frac{1}{\alpha^*(K,\bm{\Phi},\gamma)}=\mathbb{E}_{{\bm{y},\bm{\xi},\RR}}\left[\max_{\cc\in\{0,1\}^{K}}\underset{{{\bm{H}}_\cc\in \mathcal{H}^\gamma_\cc(\yy,\bm{\Sigma}|\bm{R})}}{\min}\displaystyle   \;\frac 1P\sum_{\mu=1}^P{\Vert \bm{H}^\mu_{\cc}-\xxi^\mu_\cc\Vert_2^2}\right].
\end{align}
{\bf Comparison to the context-less formula.} The capacity formula in Eq.~\eqref{eq:capacity_supmat} is a generalization of the context-less capacity $\alpha^{*}_{\cancel{C}}$ for correlated neural manifolds derived in Eq.~(5) of \cite{wakhloo2023linear}:
\begin{align}
    \frac{1}{\alpha^*_{\cancel{C}}(\gamma)}=\mathbb{E}_{{\bm{y},\bm{\xi},\RR}}\left[\underset{{{\bm{H}}\in \mathcal{A}^\gamma(\yy,\bm{\Sigma})}}{\min}\displaystyle   \;\frac 1P\sum_{\mu=1}^P{\Vert \bm{H}^\mu-\xxi^\mu\Vert_2^2}\right],
\end{align}
where the fields are independent of contexts and the constraint only depends on the manifold shapes and correlations:
\begin{align}
    \mathcal{A}^\gamma(\yy,\bm{\Sigma})=\Big\{\bm{H} \in \mathbb{R}^{ P\times (D+1)}\;: \;\forall \mu=1,\ldots,P\;,\quad\min_{\ss^\mu\in\mathcal{S}} y^\mu\sum_{\mu',i'}L^{\mu i}_{\mu' i'}H^{\mu'}_{i'}s^\mu_i\geq \gamma\Big\}\;.
\end{align}
\section{Alternative example of context assignment: prototype gating}
An alternative possible choice for the gating function is \emph{prototype gating}. In this case, inspired by prototype learning \cite{rosch1975family,sorscher2022neural}, the vector $\bm{r}_k$ defines a ``prototype'' for the $k^{\rm th}$ context. Then, the gating function $g_c$ assigns each point to the closest context. In this case, $K=|M|$. The context vectors are normalized and 
\begin{align}
g_c(\{\xx^\top\rr_k\}_{k=1}^K) =\begin{cases}
1 &\text{if} \;\;c=\arg\underset{k\in\{1,\ldots,K\}}{\min}\left\{\Vert \xx-\rr_k\Vert^2_2\right\}=\arg\underset{k\in\{1,\ldots,K\}}{\max}\left\{ \xx^\top\rr_k\right\}\\
0 &\text{otherwise}
\end{cases}.
\end{align}
In some cases, we can express half-space gating functions using prototype gating. For instance, the example considered in Sec.~IIIC of the main text is equivalent to prototype gating with two context vectors $\bm{r}_1=-\bm{r}_2$. However, we need an exponential number of context vectors to achieve equivalent expressivity with protype gating, which makes half-space gating more efficient at partitioning the input space.
We compute explicitly the capacity of prototype gating in the special case of random points in the next section.
\section{Special case: capacity for random points}
We focus on the special case of random uncorrelated points. We have that $D=0$, $\Sigma^{\mu i}_{\nu j}=\delta_{\mu\nu}\delta_{ij}$ and we can set $y^\mu=1$ without loss of generality. In particular, we are interested in half-space gating $g_\cc\left(\{\RR_k\}_{k=1}^K\right)=\prod_{k=1}^K\delta_{\Theta({\ss}^\top\RR_k),c_k}$ and prototype gating $g_c\left(\{\RR_k\}_{k=1}^K\right)=\delta_{c,\arg\underset{k}{\max}\left\{ \ss^\top\RR_k\right\}}$. The average over $\HH_\cc$ in Eq.~\eqref{eq:SRSsimplified} simplifies to
\begin{align}
\begin{split}
\frac 1P\ln\mathbb{E}_{\HH_\cc} &\left[\prod_{\mu=1}^P\Theta(J^{(\cc)}_\mathcal{S}(y^\mu\LL^\mu\bm{H}_\cc,\LL^\mu\RR)-\gamma)\right]=\\&=\frac 1P\Omega_\cc(\{R_k\}_{k=1}^K)\ln\left(\prod_{\mu=1}^P\int\frac{\dd H^\mu_{0,\cc}\,\ee^{- {\left(H^\mu_{0,\cc}-\sqrt{q_{\cc\cc}}\xi^\mu_{0,\cc}\right)^2}/{2(1-q_{\cc\cc})}}}{\sqrt{2\pi(1-q_{\cc\cc})}}\Theta\left(H^\mu_{0,\cc}-\gamma\right)\right)
\\&=\Omega_\cc(\{R_k\}_{k=1}^K)\ln\left(\frac 12 \erfc\left(\frac{\gamma-{\xi}_\cc}{\sqrt{2\,(1-q_{\cc\cc})}}\right)\right)\;,
\end{split}
\end{align}
where for simplicity we have dropped the indices $\mu$ and $0$ from $\xi_\cc\sim\mathcal{N}(0,1)$, and we have defined 
\begin{align}
    \Omega_\cc(\{R_k\}_{k=1}^K)=\begin{cases}
        \prod_{k=1}^{K}\Theta\left((-1)^{c_k}R_k\right) & \text{if}\;\; g_\cc=\text{half-space}\\
       \delta_{c,\arg\underset{k}{\max}R_k} & \text{if}\;\; g_\cc=\text{prototype}
    \end{cases}
\end{align}
It is useful to introduce the \emph{context probability} 
\begin{align}
    P(\cc|\bm{\Phi})=\mathbb{E}_{\bm{R}}\left[\Omega_\cc(\{R_k\}_{k=1}^K)\right]\in[0,1]\;,\qquad \bm{R}\sim\mathcal{N}(\bm{0},\bm{\Phi})\;.
\end{align}
The replica-symmetric action in Eq.~\eqref{eq:SRSsimplified} simplifies to
\begin{align}
    S_{\rm RS}=\frac 12 \sum_{\cc\in\{0,1\}^{K}}\left[\ln(1-q_{\cc\cc})+\frac{q_{\cc\cc}}{1-q_{\cc\cc}}\right]+\alpha \,\mathbb{E}_{{\xi}_\cc}\left[\sum_{\cc\in\{0,1\}^K}P(\cc|\bm{\Phi})\ln\left(\frac 12\erfc\left(\frac{\gamma-{\xi}_\cc}{\sqrt{2\,(1-q_{\cc\cc})}}\right)\right)\right]\;.\label{eq:SPEpoints_diagonalassumption}
\end{align}
We can define the auxiliary function
\begin{align}
    \hat S_{\rm min}=\mathbb{E}_{\xi_{\tilde\cc}}\left[\frac 12 \left(\ln(1-q_{\tilde\cc\tilde\cc})+\frac{q_{\tilde\cc\tilde\cc}}{1-q_{\tilde\cc\tilde\cc}}\right)+\alpha P(\tilde\cc|\bm{\Phi})\ln\left(\frac 12\erfc\left(\frac{\gamma-{\xi}_{\tilde\cc}}{\sqrt{2\,(1-q_{\tilde\cc\tilde\cc})}}\right)\right)\right],
\end{align}
where in this case $\tilde\cc={\rm arg}\underset{\cc\in\{0,1\}^K}{\min}P(\cc|\bm{\Phi})\ln\left(\frac 12\erfc\left(\frac{\gamma-{\xi}_\cc}{\sqrt{2\,(1-q_{\cc\cc})}}\right)\right)$. The capacity transition can be obtained  by a similar argument as for the manifold case, where
\begin{align}
    \ln\left(\frac 12\erfc\left(\frac{\gamma-{\xi}_{\tilde\cc}}{\sqrt{2\,(1-q_{\tilde\cc\tilde\cc})}}\right)\right)\underset{q_{\tilde\cc\tilde\cc\rightarrow 1}}{\simeq}-\frac{\left(\gamma-{\xi}_{\tilde\cc}\right)^2}{2(1-q_{\tilde\cc\tilde\cc})}\Theta(\gamma-\xi_{\tilde\cc})\;.
\end{align}
\subsection{Uniform off-diagonal context correlations.}
\label{appendix:uniform_correlations}
We consider a special case of context correlation matrix: $\bm{\Phi}=(1-\phi)\bm{I}_K + \phi \bm{1}_K\bm{1}_K^\top$, where $\bm{1}_K$ denotes the $K-$dimensional vector with all entries equal to one. Thus, we have that $\det\bm{\Phi}=(1-\phi)^{K-1}\left(1+(K-1)\phi\right)$ and $\bm{\Phi}^{-1}=(1-\phi)^{-1}\left(\bm{I}_K - \phi\bm{1}_K\bm{1}_K^\top/(1+\phi(K-1))\right)$ by Sherman-Morrison lemma.

In the case of half-space gating, the context probability reduces to
\begin{align}
\begin{split}
    \label{eq:context_proba_points}P(\cc|\phi)&=\int \left[\prod_{k=1}^K\frac{\dd R_k}{\sqrt{2\pi(1-\phi)}}\,\Theta\left((-1)^{c_k}R_k\right)\right]\sqrt{\frac{1-\phi}{1+(K-1)\phi}}\ee^{-\frac{\bm{R}^\top\bm{R}}{2(1-\phi)}+\frac{\phi(\bm{R}^\top\bm{1}_K)^2}{2(1-\phi)\left(1+(K-1)\phi\right)}}\\
    &=\sqrt{\frac{1-\phi}{1+(K-1)\phi}}\mathbb{E}_\eta\left[\prod_{k=1}^K\int\frac{\dd R_k}{\sqrt{2\pi(1-\phi)}}\,\Theta\left((-1)^{c_k}R_k\right)\ee^{-\frac{R_k^2}{2(1-\phi)}+\frac{\sqrt{\phi}\eta R_k}{\sqrt{(1-\phi)(1+(K-1)\phi)}}}\right]\\
&=\mathbb{E}_{\eta}\left[2^{-K}\prod_{k=1}^K\left[1+(1-2c_k) \erf\left(\frac{\sqrt{\phi}\eta}{\sqrt{2(1-\phi)}}\right)\right]\right]
\end{split}
\end{align}
where $\eta\sim\mathcal{N}(0,1)$.
In the case of prototype gating, the context probability is simply uniform
\begin{align}
    P(c|\phi)=\frac{1}{K},
\end{align}
by symmetry, because all off diagonal elements of the covariance are equal. Therefore, for random points with this simple correlation matrix, the capacity for prototype gating is $\alpha^*=2K/(1+\gamma^2)$, in agreement with the seminal result by Cover \cite{cover1965geometrical}.   

\section{Details on the  numerical estimation of replica formula}
In the numerical experiments, a manifold is modeled as a point cloud. For simplicity, we focus on the case where every manifold has the same number of points. Let $N$ be the number of units, let $P$ be the number of manifolds, and let $M$ be the number of points per manifold. The $\mu$-th manifold $\cM^\mu$ consists of a collection of points $\{x^\mu_1,\dots,x^\mu_M\}$ where $x^\mu_i\in\Real^N$ for each $i=1,\dots,M$. Without loss of generality, we assume points in the same manifold are linearly independent. 

The algorithm that computes capacity consists of two steps: (1) compute the correlations between manifolds, (2) sample the quenched disorder and calculate the capacity. The following are the pseudocodes for each step.
In the following algorithms, $A\gets\{x_i\}$ denotes putting the collection of vectors $\{x_i\}$ on the rows of $A$.\\\\
{\bf Step 1: Compute the correlations between manifolds.}
In this step we first compute the center of mass of each manifold $x_0^\mu$ and center each point as $\bar{x}_i^\mu=x_i^\mu-x_0^\mu$. Next, we find a basis for the centered points for each manifold and estimate the covariance tensor $C^{\mu,i}_{\nu,j}$ accordingly. This step is exactly the same as the corresponding step in~\cite{wakhloo2023linear,chou2024}. See Algorithm 1 for details.
\begin{table}[ht]
    \renewcommand{\arraystretch}{1.2} 
    \noindent \textbf{Algorithm 1:} Compute the correlations between manifolds

    \vspace{2mm}
    \begin{tcolorbox}[colframe=black!60, colback=white, sharp corners]

    \begin{flushleft}
    \textbf{Input:} \\
        $\{\mathcal{M}^\mu\}_{\mu=1}^P$: $P$ manifolds of $M$ points in an ambient dimension of $N$. \\
    \end{flushleft}

    \begin{flushleft}
     \textbf{Output:} \\
        $C$: Correlation tensor of shape $P\times(M+1)\times P\times(M+1)$. \\
        $\{\widetilde{\mathcal{M}}^\mu\}_{\mu=1}^P$: $P$ manifolds of $M$ points in an ambient dimension of $M+1$. \\
    \end{flushleft}

    \begin{flushleft}
    \begin{tabular}{r l} 
    1:  & \textbf{for} $\mu = 1$ to $P$ \textbf{do} \\
    2:  & $\quad x_0^\mu \gets \frac{1}{M} \sum_i x_i^\mu$. \\
    3:  & $\quad \bar{x}_i^\mu \gets x_i^\mu - x_0^\mu$, for each $i=1,\dots,M$. \\
    4:  & \textbf{end for} \\
    
    5:  & \textbf{for} $\mu = 1$ to $P$ \textbf{do} \\
    6:  & $\quad Q \gets$ an orthogonal basis for $\textsf{span}(\{\bar{x}_i^\mu\})$. \\
    7:  & $\quad z^\mu_i \gets Q^\top \bar{x}^\mu_i$, for each $i=1,\dots,M$. \\
    8:  & $\quad \widetilde{\mathcal{M}}^\mu \gets \{\tilde{x}^\mu_i\}_{i=1}^M$, where $\tilde{x}^\mu_i = [z^\mu_i\ 1]$. \\
    9:  & $\quad q_0^\mu \gets \bar{x}_0^\mu$. \\
    10: & $\quad q_i^\mu \gets Q[i]$, for $i=1,\dots,M$. \\
    11: & \textbf{end for} \\

    12: & Let $C$ be the $P\times(M+1)\times P\times(M+1)$ tensor with $ C^{\mu,i}_{\nu,j} = \langle q^\mu_i, q^\nu_j \rangle$, \\
       & for each $\mu, \nu \in \{1, \dots, P\}$ and $i, j \in \{0, 1, \dots, M\}$. \\

    13: & \textbf{return} $C, \{\widetilde{\mathcal{M}}^\mu\}_{\mu=1}^P$. \\
    \end{tabular}
    \end{flushleft}

    \end{tcolorbox}

\end{table}
{\bf Step 2: Sample the quenched disorder and calculate capacity.}
In this step, we first sample $K$ context hyperplanes $\{\mathbf{R}_k\}_{k=1}^K$. Next, we conduct $n_t$ ($n_t=100$ in all experiments) repetition of random sampling $\xi,\{R_k\}$ for averaging. For each repetition, for each context we sample the manifold disorder and the create quadratic programming problem with the points that lie in the context (as defined by $\{R_k\}_{k=1}^K$). Finally, capacity is the inverse of the average of the maximum over the outcome of the quadratic programming problems. Note that when $K=0$ this step is exactly the same as the corresponding step in~\cite{wakhloo2023linear,chou2024}. See Algorithm 2 for details.


\begin{table}[ht]
    \renewcommand{\arraystretch}{1.2}
    \noindent \textbf{Algorithm 2:} Sample the quenched disorder and calculate capacity

    \vspace{2mm}
    \begin{tcolorbox}[colframe=black!60, colback=white, sharp corners]
    
    \begin{flushleft}
     \textbf{Input:} \\
        $C$: Correlation tensor of shape $P\times(M+1)\times P\times(M+1)$  \\
        $\{\widetilde{\mathcal{M}}^\mu\}_{\mu=1}^P$: $P$ manifolds of $M$ points in an ambient dimension of $M+1$  \\
        $n_t$: number of samples  \\
        $\kappa$: margin  \\
        $K$: number of context hyperplanes  \\
        $\psi$: correlation between a pair of context hyperplanes

    \vspace{2mm}
    \textbf{Output:} $\alpha$: capacity
  \end{flushleft}
    \begin{flushleft}
    \begin{tabular}{r l}
    1:  & Sample $\{\mathbf{R}_k\}_{k=1}^K$, where $\mathbf{R}_k \in \mathbb{R}^{P\times(M+1)}$  and the $i$-th entry of $\mathbf{R}_k$ is sampled from a multivariate Gaussian\\
       & distribution with mean $0$ and covariance $(1-\psi)\delta_{kk'}+\psi$. \\

    2: & $\quad \textsf{conv}(\{\widetilde{\mathcal{M}}^\mu\}) \leftarrow \bigcup_\mu \bigcup_k \left\{ y + \frac{\mathbf{R}_k^\top y}{\mathbf{R}_k^\top(x - y)}(x - y)  : x,y \in \widetilde{\mathcal{M}}^\mu, \mathbf{R}_k^\top x \geq 0, \mathbf{R}_k^\top y \leq 0 \right\}$. \\

    3:  & \textbf{for} $\ell = 1$ to $n_t$ \textbf{do} \\
    4:  & $\quad$ \textbf{for } $b \in \{-1,1\}^K$ \textbf{do} \\
    5:  & $\quad\quad$ $T_k \; \leftarrow$ a vector sampled from an isotropic Gaussian distribution in $\mathbb{R}^{P\times(M+1)}$. \\
    6:  & $\quad\quad$ $y\;\leftarrow$ a random vector from $\{-1,1\}^P$. \\
    7:  & $\quad\quad$ $L \leftarrow \textsf{Cholesky}(C)$ \hfill \textit{(Cholesky decomposition)} \\
    8:  & $\quad\quad$ $A \leftarrow I_{P(M+1)}$. \\
    9:  & $\quad\quad$ $q \leftarrow -T_k$. \\
    10: & $\quad\quad$ $G \leftarrow (y \odot L)  \textsf{conv}(\{\widetilde{\mathcal{M}}^\mu\})$. \\
    11: & $\quad\quad$  $h \leftarrow \kappa \cdot \mathbf{1}_{P(M+1)}$. \\
    12: &  $\quad\quad \textsf{output} \leftarrow qp(A, q, G, h).$ \hfill \textit{(Quadratic programming with linear constraints)}\\
    13: & $\quad\quad$ $\textsf{norm}^b_\ell \leftarrow \textsf{output}[\text{``value''}]$. \\
    14: & $\quad$\textbf{end for}\\
    15: & $\quad$ $\textsf{norm}_\ell \leftarrow \max_b \{\textsf{norm}^b_\ell\}$. \\
    16: & \textbf{end for}\\
     17: & 
        $ \alpha \leftarrow {1}/\left({\frac{1}{n_t} \sum_{\ell} \textsf{norm}_\ell}\right)$. \\
    18: & \textbf{return} $\alpha$. \\
    \end{tabular}
    \end{flushleft}

    \end{tcolorbox}

\end{table}
\begin{figure}
    \centering
    \includegraphics[scale=0.65]{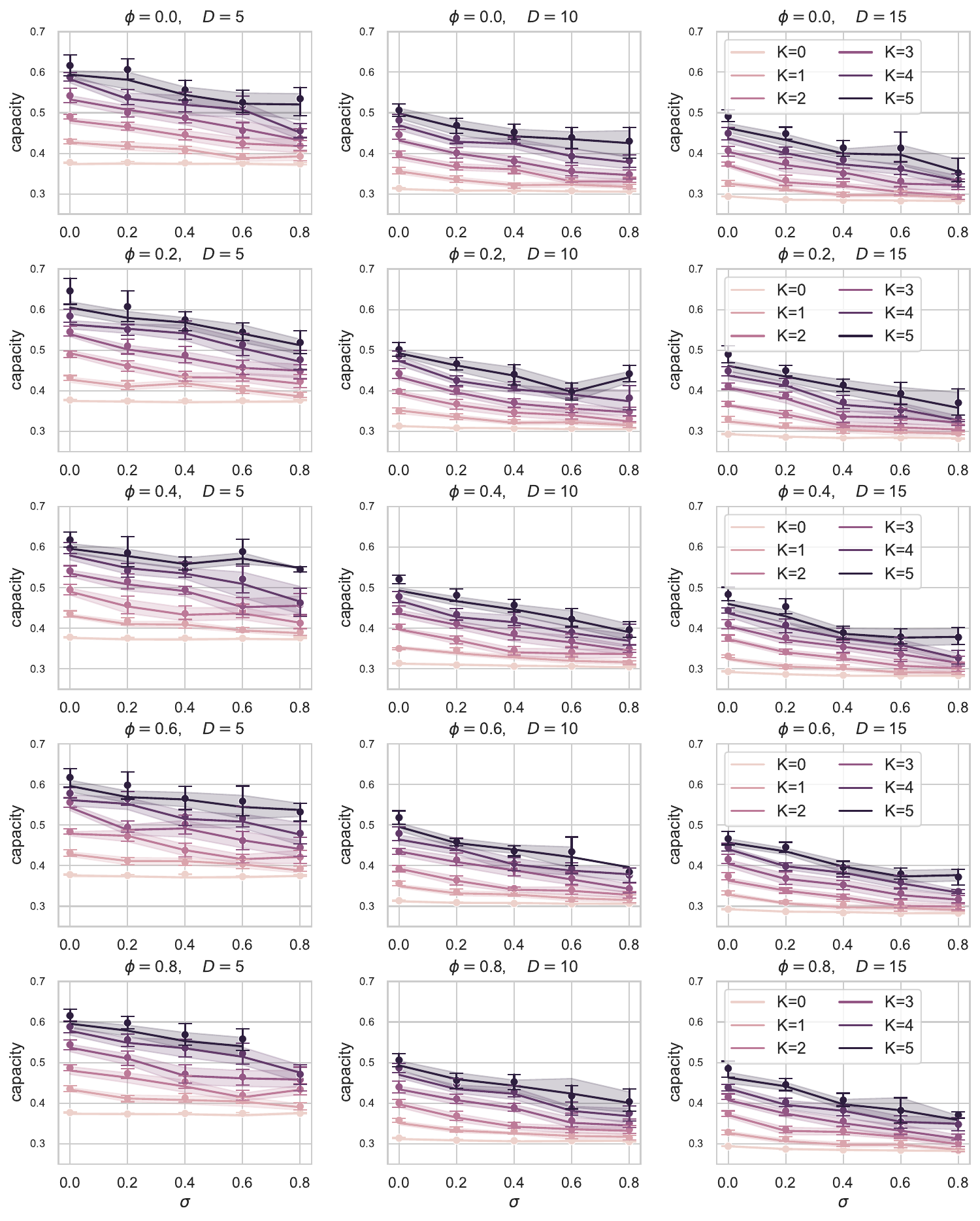}
    \caption{Capacity as a function of the manifold correlation, that for visibility purposes we take uniform $\Sigma^{\mu i}_{\nu j}=\sigma$ for all pairs of manifolds $(\mu,\nu)$ and directions $(i,j)$ for synthetic spherical manifolds. Subplot in different rows represent different values of context correlation $\phi\in [0,0.2,0.4,0.6,0.8]$, while different columns represent different latent dimension $D\in [5,10,15]$, embedded in ambient dimension $N=4000$. We take $P=50$ spherical manifolds, each with $M=50$ points. Each panel depicts the capacity for $2^K=1,2,4,8,16,32$ contexts, represented by different colors. Full lines mark theoretical predictions while dots mark simulations.}
    \label{fig:synthetic_extended}
\end{figure}
\section{Details on the  numerical experiments}
For the numerical checks in Fig.~4 of the main text and Fig.~\ref{fig:synthetic_extended} below, we adopt the data generating process as described in~\cite{chou2024} (Fig.~SI2). Specifically, we consider $N=4000$, $P=50$, and $M=50$. Following the convention in previous work~\cite{chung2018classification,stephenson2021geometry,wakhloo2023linear,chou2024}, we conduct numerical check by using ``simulated capacity'' as the ground truth and comparing it with the result from out replica formula. Concretely, the simulated capacity is estimated by binary searching the number of neurons for the critical $N^*$. We use the code from~\cite{chung2018classification} for all our experiments.

For the imagenet experiment in Fig.~7 of the main text, we adopt the setting as described in~\cite{cohen2020separability,wakhloo2023linear,chou2024}. Specifically, the neural responses are extracted from a pretrained ResNet-50 architecture trained on ImageNet with the supervised learning algorithm. We focus on 7 layers in ResNet-50: \textsf{x}, \textsf{relu}, \textsf{layer1.1.relu\_2}, \textsf{layer2.3.relu\_2}, \textsf{layer3.5.relu\_2}, \textsf{layer4.2.relu\_2}, \textsf{avgpool}. For each random repetition, we fix a random projection matrix for each layer and project the neural activations to a $2000$ dimensional subspace.
In each repetition, we randomly select $40$ categories from the $1000$ ImageNet categories and randomly select $40$ images from the top 10\% accurate images of each category to for the manifolds (we follow the same protocol as in ref.~\cite{cohen2020separability}).

Finally, we remark on the difficulties in terms of numerically matching the simulated capacity and replica formula. The central challenge lies in the fact that the number of points per manifold grows exponentially in $K$ (line 2 in Algorithm 2). As the running time of the algorithm for both replica formula and simulated capacity depend on the number of points per manifold, it becomes computationally expensive and time-consuming for $K$ larger than $3$. Furthermore, the required ambient dimension $N$ for more accurate estimations for capacity also scales linearly with the number of points per manifold. And for simulated capacity, the required number of manifolds $P$ also needs to scale up accordingly. Meanwhile, the algorithm for replica formula does not require the scale-up of $P$, hence providing a computational advantage over the simulated capacity.

\end{document}

%% file: commands.tex
\newcommand{\xx}{{\bm{x}}}

\newcommand{\UU}{{\bm{U}}}
\renewcommand{\ss}{{\bm{s}}}
\newcommand{\yy}{{\bm{y}}}
\newcommand{\LL}{{\bm{L}}}
\newcommand{\cc}{{\bm{c}}}
\newcommand{\uu}{{\bm{u}}}
\newcommand{\xxi}{{\bm{\xi}}}
\newcommand{\ww}{{\bm{w}}}

\newcommand{\HH}{{\bm{H}}}
\newcommand{\rr}{{\bm{r}}}

\newcommand{\ee}{{\rm e}}

\newcommand{\RR}{\bm{R}}

\newcommand{\erfc}{{\rm erfc}}
\newcommand{\sign}{{\rm sign}}

\usepackage{bbm}
\usepackage{bm}
\usepackage{mathtools}
\usepackage{cancel}